\documentclass[useAMS,usenatbib]{mnras}

\usepackage{aas_macros}
\usepackage{hhline}
\usepackage{bmpsize}
\usepackage{amssymb,amsmath,amsfonts,bm}
\usepackage{comment}
\usepackage{pdflscape}
\makeatletter
\renewcommand\PLS@Rotate[1]{}
\makeatother
\usepackage{afterpage}
\usepackage{multirow}
\usepackage{xcolor}
\usepackage[caption=false]{subfig}
\usepackage{graphicx}
\usepackage{lipsum}
\usepackage{hyperref}
    \hypersetup{
        colorlinks,
        linkcolor={red!80!black},
        citecolor={blue},
        urlcolor={green!80!black}
    }

\graphicspath{.{/mnt/data/Work/Clusters/paper/figs/}}
\def\mjyb{mJy beam$^{-1}$ }
\def\mjyam{mJy arcmin$^{-2}$ }

\def\RRI{$^{1}$}
\def\Victoria{$^{2}$}
\def\Leiden{$^{3}$}
\def\Curtin{$^{4}$}
\def\CASS{$^{5}$}
\def\CAASTRO{$^{6}$}
\def\USydney{$^{7}$}
\def\UWA{$^{8}$}
\def\UHertfordshire{$^{9}$}
\def\UMelbourne{$^{10}$}
\def\ASTRON{$^{11}$}

\title[MWA Observations of merging galaxy clusters]
{A study of halo and relic radio emission in merging clusters
using the Murchison Widefield Array}

\author[George et. al.]{
L.~T.~George\RRI\thanks{lijo@rri.res.in}, 
K.~S.~Dwarakanath\RRI\thanks{dwaraka@rri.res.in}, 
M.~Johnston-Hollitt\Victoria, 
H.~T.~Intema\Leiden,
\newauthor{
N.~Hurley-Walker\Curtin, 
M.~E.~Bell\CASS$^,$\CAASTRO, 
J.~R.~Callingham\CASS$^,$\CAASTRO$^,$\USydney, 
Bi-Qing For\UWA, 
B.~Gaensler\USydney,
}
\newauthor{
P.~J.~Hancock\Curtin$^,$\CAASTRO,
L.~Hindson\UHertfordshire,
A.~D.~Kapi\'nska\CAASTRO$^,$\UWA, 
E.~Lenc\USydney, 
}
\newauthor{
B.~McKinley\UMelbourne, 
J.~Morgan\Curtin, 
A.~Offringa\ASTRON, 
P.~Procopio\UMelbourne, 
L.~Staveley-Smith\UWA, 
}
\newauthor{
R.~B.~Wayth\Curtin$^,$\CAASTRO, 
Chen~Wu\UWA, 
Q.~Zheng\Victoria, 
}
\\
$^{1}$Raman Research Institute, Bangalore 560080, India\\
$^{2}$School of Chemical \& Physical Sciences, Victoria University of Wellington, PO Box 600, Wellington 6140, New Zealand\\
$^{3}$Leiden University, PO Box 9500, 2300 RA Leiden\\
$^{4}$International Centre for Radio Astronomy Research, Curtin University, Bentley, WA 6102, Australia\\
$^{5}$CSIRO Astronomy and Space Science (CASS), PO Box 76, Epping, NSW 1710, Australia\\
$^{6}$ARC Centre of Excellence for All-sky Astrophysics (CAASTRO), 44 Rosehill Street, Redfern, NSW 2016, Australia\\
$^{7}$Sydney Institute for Astronomy, School of Physics, The University of Sydney, NSW 2006, Australia\\
$^{8}$International Centre for Radio Astronomy Research, University of Western Australia, Crawley, WA 6009, Australia\\
$^{9}$Centre for Astrophysics Research, School of Physics, Astronomy and Mathematics,\\
\hspace{0.3em} University of Hertfordshire, College Lane, Hatfield AL10 9AB, UK\\
$^{10}$School of Physics, The University of Melbourne, Parkville, VIC 3010, Australia\\
$^{11}$ASTRON, 7990 AA Dwingeloo, The Netherlands\\
}

\pagerange{\pageref{firstpage}--\pageref{lastpage}} \pubyear{2016}

\begin{document}

\label{firstpage}

\maketitle
	
\begin{abstract}
    We have studied radio haloes and relics in nine merging galaxy clusters
    using the Murchison Widefield Array (MWA).
    The images used for this study were obtained from the
    GaLactic and Extragalactic All-sky MWA (GLEAM) Survey
    which was carried out at 5 frequencies, viz. 88, 118, 154, 188 and 215 MHz.
    We detect diffuse radio emission in 8 of these clusters.
    We have estimated the spectra of haloes and relics in these clusters 
    over the frequency range $80-1400$ MHz; the first such attempt to estimate
    their spectra at low frequencies.
    The spectra follow a power law 
    with a mean value of $\alpha = -1.13\pm0.21$ for haloes and 
    $\alpha = -1.2\pm0.19$ for relics where, $S \propto \nu^{\alpha}$. 
    We reclassify two of the cluster sources as radio galaxies.
    The low frequency spectra are thus an independent
    means of confirming the nature of cluster sources.
    Five of the nine clusters host radio haloes.
    For the remaining four clusters, we place upper limits on the
    radio powers of possible haloes in them.
    These upper limits are a factor of $2-20$ below those expected
    from the $L_{\rm X}-P_{\rm 1.4}$ relation.
    These limits are the lowest ever obtained and
    the implications of these limits to the hadronic model of halo emission
    are discussed.
\end{abstract}

\newpage


\section{Introduction}
    According to the hierarchical model for structure formation, galaxy clusters 
    are formed as a result of several successive mergers between smaller sub-clusters. 
    These mergers are extremely powerful events that can release an enormous
    amount of energy ($\sim 10^{63}-10^{64}$ ergs, \citealt{hoeft08}) 
    into the surrounding Intra-Cluster Medium (ICM). A fraction of this energy
    could be used to amplify magnetic fields in clusters \citep{carilli02,subra06}
    and/or accelerate relativistic electrons in the ICM \citep{petrosian01,brunetti01}
    as a result of which synchrotron radiation could be produced in the ICM.

    This extended ($\sim$ Mpc), diffuse ($\sim$ few \mjyam at 1.4 GHz) radio emission 
    from galaxy clusters which is not associated with any galaxy or compact object, 
    but rather from the ICM gas itself comes in two forms -- {\it radio haloes} 
    and {\it radio relics} \citep{feretti12,brujones14}.
    Radio haloes are found in the central regions of clusters while the relics 
    are found mostly near the peripheries of clusters,
    towards the edges of the X-ray emission.
    Haloes usually have a smoother, circular  morphology while relics are usually 
    elongated and arc-like.

    Clusters which host these haloes and relics are rare in the universe.
    Below a redshift of $z=0.2$, only $\sim 5\%$ of all galaxy clusters
    host these objects \citep{giovannini99}.
    In a study of highly X-ray luminous clusters
    ($L_{\rm X}\gtrsim 10^{45}$ ergs/s) with $0.2<z<0.4$, \citep{venturi07,venturi08}
    (and its follow-up by \cite{kale13})
    found that $38\%$ of galaxy clusters hosted radio haloes.
    Additionally, highly X-ray luminous galaxy clusters are also expected to have
    higher radio powers as well
    \citep{liang2000,bacchi03,cassano06,brunetti07,brunetti09,rudnick09}.
    However, most previous efforts to increase the number of haloes and relics
    have concentrated on X-ray luminous clusters and there has yet to be a
    completely unbiased search for diffuse cluster emission.


    
    Relics are believed to trace the outward going shocks produced at the time of
    cluster merger.
    The theory that explains this mechanism, 
    Diffusive Shock Acceleration (DSA) theory \citep{blandford87,jones91,ensslin98}
    suggests 
    that electrons in the ICM suffer multiple collisions across the two fronts of the 
    outward going shock as a result of which they get accelerated diffusively.
    Clusters with double relics i.e., relics which are located on opposite ends of 
    the cluster centre and trace the outgoing shocks are 
    the best examples in support of this theory.
    However, recent analysis \citep{vazza14,vazza15,vazza16}
    suggests that this might not be completely true.
    There are still large uncertainties on the mechanism by which the electrons
    can be accelerated or reaccelerated by such low Mach number shocks
    as those observed in clusters.

    The origin of radio haloes is also not well understood. The basic problem with
    the existence of Mpc-sized radio haloes is that the synchrotron life times of
    the radiating electrons is about ten times smaller than their diffusion
    time scales over the sizes of radio haloes. A consequence of this is that 
    relativistic electrons can not be produced in the active galaxies in clusters
    and transported all over the cluster. The existence of Mpc-sized haloes
    appears to imply {\it in-situ} acceleration.
    The most popular model used to explain them is that of {\it turbulent acceleration} 
    or the {\it primary} model \citep{petrosian01,brunetti01}.
    According to this model, the electrons in the ICM are accelerated as a result of 
    turbulence which is generated in the aftermath of a cluster merger.
    A 'proof' of such a scenario is the fact that clusters with disturbed morphologies
    are known to host radio haloes than relaxed clusters \citep{buote01,cassano10b}.
    An alternative to this model
    is the {\it hadronic} or {\it secondary} model \citep{dennison80,blacola99,pfrosslin04}
    according to which ineleastic collisions between relativistic protons and thermal
    protons in the ICM produces pions which decay to produce electrons and gamma rays.
    The lack of gamma ray detections in galaxy clusters 
    \citep{aharonian09a,aharonian09b,ackermann10,aleksic10}
    is a major shortcoming of this model.
    It is also claimed that the synchrotron emission from the secondary model
    is a factor of 10 lower than that from the primary model.


    Radio haloes are known to exhibit an empirical correlation between
    the radio power of the halo at 1.4 GHz
    and the X-ray luminosity of the galaxy cluster that hosts the halo
    \citep{liang2000,feretti00,govoni01,bacchi03,brunetti07,brunetti09}.
    Recent GMRT (Giant Metrewave Radio Telescope) observations of
    galaxy clusters \citep{venturi07,venturi08} also found 
    that galaxy clusters show a \emph{bi-modal} nature
    in the $L_{\rm X}-P_{\rm 1.4}$ plot.
    In many clusters where no radio halo was detected,
    upper limits to the halo emission were placed 
    that are a factor of 2--3 below the expected radio power.
    While it is possible that these clusters do not contain any radio haloes at all, 
    it could also just be that at the sensitivity limits of the
    current generation of radio telescopes (e.g. GMRT, JVLA (Jansky Very Large Array))
    any possible weak halo emission from clusters is not detectable. 
    However, if the hadronic model is to be believed
    then there will always be a component of diffuse radio emission in galaxy clusters
    due to relativistic protons.

    For this reason it is important to study non-detections of radio haloes just as much as detections.
    Using next generation telescopes like the MWA \citep{lonsdale09,bowman13,tingay13}, LOFAR 
    (LOw Frequency ARray, \citealt{vanhaarlem13}) and the upcoming SKA (Square Kilometer Array, \citealt{dewdney13}), 
    which have better low surface brightness sensitivity and $uv$-coverage at short baselines 
    as compared to existing telescopes, it could be possible to detect previously undetected radio haloes.
    The radio power in these haloes will also tell us what fraction of the power in haloes 
    is contributed due to the hadronic model as compared to the turbulence model.

    With this in mind, we decided to observe merging galaxy clusters with the MWA.
    These clusters were chosen from literature on the basis of their position in the sky (Southern Hemisphere)
    as well as existing observations of the cluster and whether or not any haloes and/or relics 
    were detected in them.
    Based on the above criteria we came up with a list of 9 galaxy clusters all of which
    are claimed to host a halo and / or a relic based on higher frequency (1.4 GHz) observations.
    These clusters are -- Abell 13, Abell 548b, Abell 2063, Abell 2163,
    Abell 2254, Abell 2345, Abell 2744, PLCK G287.0+32.9 and RXC J1314.4-2515.

    In section 2,
    we give the details of the observations of these clusters made with the MWA 
    as well as with the GMRT.
    The primary results of the paper are given in Section 3, with a 
    a discussion of these results in Section 4. The main conclusions of the
paper are summarized in Section 5.
    The cosmology used in this paper is as follows:
    $\Omega_0=0.3, \Omega_\Lambda=0.7, H_0 = 68$ km/s/Mpc

\section{Observations and Analysis}

	\subsection{GLEAM Survey Images}
    The clusters studied in this paper were observed as part of the
    GaLactic and Extragalactic All-sky MWA (GLEAM) Survey \citep{wayth15}. 
    The survey was carried out at 5 frequency bands between 72 and 230 MHz,
    centred on 88, 118, 154, 188 and 215 MHz with 30.72 MHz bandwidth each.
    Each full band was further divided into four sub-bands of 7.68 MHz bandwidth.
    The GLEAM Survey was carried out over two years.
    During the first year,
    the frequency resolution of the survey was 40 kHz
    while the time resolution was 0.5s.
    In the second year, the frequency and time resolutions
    were changed to 10 kHz and 2s, respectively.
    For our purposes we used the year one data.

    The survey was carried out at seven declination settings
    ($\delta = +18.6^{\circ}, +1.6^{\circ},-13.0^{\circ}, -26.7^{\circ}, -40.0^{\circ}, -55.0^{\circ}, -72^{\circ}$),
    utilizing a drift-scan method.
    At the beginning of the observation the telescope was electronically set to
    one of the seven declinations
    and measurements were taken sequentially looping over the five frequencies
    every two minutes (112 sec) as the sky drifted overhead.
    A set of calibrators were observed throughout each of the observing nights.

    This raw data was then analysed and sent through a pipeline to produce the final images used in this paper.
    The basic steps of the analysis are as follows.
    
    For each scan:
        \begin{itemize}
            \item{A single bright source was used for a first-pass calibration on all the observations \citep{nhw14};}
            \item{Cleaning of the images was performed using \textsc{WSClean} \citep{offringa14};}
            \item{The primary beam model for the MWA, as described by \cite{sutinjo15}, was used to transform to astronomical Stokes;}
            \item{Assuming the sky to be unpolarized, Stokes Q, U and V were set to zero and, using the same beam model, converted back to instrumental Stokes;}
            \item{Self-calibration was now performed using this new sky model to produce the final multi-frequency synthesis images;}
        \end{itemize}

    The GLEAM images were calibrated in 3 steps:
    first, at each frequency, the model of a bright source
    was used to transfer the complex antenna gains to the entire drift scan data
    of the night.
    Second, self-calibration was performed as described earlier.
    Finally, bright point sources ($>8\sigma$) were chosen and
    cross-matched with the VLA Lowfrequency Sky Survey Redux (VLSSr) at 74 MHz, 
    Molonglo Radio Catalog (MRC) at 408 MHz and NRAO VLA Sky Survey (NVSS) at 1400 MHz.
    A power law was fit to the spectra of the point sources and
    based on their expected to observed flux densities,
    a declination dependant average scaling factor
    was estimated at every MWA frequency and applied to the images.
    All the snapshots obtained during a night's observations
    were then combined in an inverse-noise-weighted fashion
    to produce mosaics at every frequency.
    Note that during this procedure,
    the absolute flux density scale of sources was set to
    the Baars scale \citep{baars77}.
    Details on all the above procedures can be found in \cite{nhw17}.
    For most of the sources used in this paper,
    the uncertainties in their flux density measurements is $\sim 8\%$.
    The images used in this paper are all taken from the GLEAM survey.
    
    In this paper we make use of the 30.72 MHz bandwidth images
    centred at 88, 118 and 154 MHz in addition to the 60 MHz bandwidth
    wideband image centred at 200 MHz.   
    These wideband images were made using observations in the frequency range
    $170-–231$ MHz.
    These wideband images compromise between improved sensitivity and resolution
    and represent the best images to search for diffuse cluster emission. These
    wideband images have a resolution of $\sim2\arcmin$
    and an RMS (Root Mean Square) value of $\sim6$ \mjyb at 200 MHz.
        
    Estimation of RMS values in the GLEAM images was carried out using
    the software package Background And Noise Estimation (BANE) written by Paul Hancock
    \footnote{\url{https://github.com/PaulHancock/Aegean/wiki/BANE}}.
    The standard method of estimating RMS from an image would be
    to estimate the mean and standard deviation in a fixed size box
    around every pixel in the image and then average it.
    However, this method is extremely time consuming
    and will also be biased due to the presence of sources in the image.

    BANE uses a slightly modified version of this algorithm
    to quickly and accurately estimate the RMS of an image.
    The software works on the principle that there is
    a high degree of correlation between adjacent pixels in a radio image.
    As such, it is not necessary to estimate the mean
    and standard deviation in a box at every pixel.
    Instead, boxes are drawn around every $N^{\rm th}$ pixel and,
    first, contribution from the source pixels is removed
    by masking pixels greater than $3\sigma$.
    This sigma clipping is performed three times and then,
    instead of the mean, the median is estimated for each grid
    and interpolated to produce the background image.
    The same process is repeated on the background subtracted image
    (data--background) and then the standard deviation of the image
    is estimated.

	\subsection{TGSS Images}
    The TGSS\footnote{TIFR (Tata Institute of Fundamental Research) GMRT Sky Survey;
    see \url{http://tgss.ncra.tifr.res.in/}}
    is a fully observed survey
    of the radio sky at 150 MHz as visible from the GMRT, 
    covering the full declination range of -55 to +90~degrees. 
    Data was recorded in half polarization (RR,LL) every 2~seconds in 256~frequency channels across 16 MHz of bandwidth (140--156~MHz). 
    Each pointing was observed for about 15 minutes, split over 3 or more scans spaced out in time to improve UV-coverage. 
    As a service to the community,
    this archival data has been processed with a fully automated pipeline
    based on the Source Peeling and Atmospheric Modelling (SPAM) package \citep{intema09,intema14}, 
    which includes direction-dependent calibration,
    modelling and imaging to suppress mainly ionospheric phase errors. 
    
    In summary, the pipeline consists of two parts: 
    a \emph{pre-processing} part that converts the raw data
    from individual observing sessions into pre-calibrated visibility data sets
    for all observed pointings, 
    and a \emph{main pipeline} part that converts
    pre-calibrated visibility data per pointing into stokes I continuum images. 
    The flux density scale is set by calibration on
    3C48, 3C147 and 3C286 using the models from \citet{scaife12}. 
    More details on the processing pipeline
    and characteristics of the data products can be found in the article on
    the first TGSS alternative data release (ADR1; \citealt{intema16}). 
    For this study,
    ADR1 images were used to create mosaics at the cluster positions. 
    These images have a resolution of $\sim25\arcsec$
    and an RMS of $\sim 5$ \mjyb at 150 MHz.

    The primary purpose of using the TGSS images is that
    since they have better resolution than the GLEAM images
    it would be easier to detect any blending of unrelated sources
    that might occur with the haloes and relics.
    Such sources could then be identified and their flux densities subtracted
    in order to accurately estimate the flux densities of the haloes and relics.

\section{Results}

    \begin{table*}
        \centering
        \begin{tabular}{l*{7}{c}}
            \hline
            Cluster                 &	\multicolumn{4}{c}{MWA ($''\times'',^{\circ}/$\mjyb)}	                &	\multicolumn{2}{c}{TGSS (\mjyb)}	\\
                                    &	88 MHz              &	118	MHz             &	154	MHz                 &	200	MHz             &   150 (25$''$)    & 150 (60$''$)	    \\ 
            \hline
            \multirow{2}{*}{A13}    &   287$\times$263, -74	&	202$\times$193, -76	&	153$\times$147, -77		&	127$\times$128, -83  \\ 
           	    					&   37.7	            &	19.3	            &	12.7                    &	7.8                 &   5.0 & 11.4	            \\ \hline 
            \multirow{2}{*}{A548b}  &   282$\times$265, 67	&	201$\times$193, 72	&	153$\times$148, 62		&	128$\times$123, 39  \\  
        	    					&   38.2 	            &	16.6	            &	12.1 	                &	6.0                &   4.5 & 9.4               \\ \hline
            \multirow{2}{*}{A2063}  &   324$\times$383, 0	&	236$\times$204, 4	&	183$\times$158, 1		&	146$\times$127, 2   \\ 
        	    					&   77.9	            &	45.3	            &	22.2	                &	19.6                &   7.1 & 17.0              \\ \hline
            \multirow{2}{*}{A2163}  &   286$\times$272, 0   &	206$\times$197, -14	&	158$\times$150, -8		&	129$\times$122, -10  \\  
        	    					&   62.8	            &	36.6	            &	20.4	                &	14.3                &   3.8 & 8.9               \\ \hline
            \multirow{2}{*}{A2254}  &   347$\times$278, 0   &   257$\times$200, 0   &	203$\times$157, 0	    &   161$\times$126, -2   \\ 
        	    					&   106.9               &	65.9                &	35.7                    &	37.6                &   4.4 & 10.9              \\ \hline
            \multirow{2}{*}{A2345}  &   293$\times$282      &	203$\times$198, -51	&	155$\times$150, -45	    &	130$\times$125, -59   \\    
        	    					&   46.4		        &	22.2	            &	16.7 	                &	7.4                &   5.0 & 12.3              \\ \hline
            \multirow{2}{*}{A2744}  &   287$\times$263, -56	&	204$\times$191, -50	&	156$\times$146, -55		&	129$\times$122, -41  \\
        	    					&   33.5	            &	16.9	            &	12.2 	                &	7.2                &   4.3 & 10.2              \\ \hline
		    \multirow{2}{*}{PLCK G287.0+32.9}  & 278$\times$264, 72	& 201$\times$194, 73 &	152$\times$147, 37 &	125$\times$122, 14 \\
        	    					&   37.9	            &	19.7	            &	12.2	                    &	5.6                &   5.0 & 10.2              \\ \hline           
            \multirow{2}{*}{RXC J1314.4-2515}  & 277$\times$264, 70	& 202$\times$194, 64 &	152$\times$149, 50	&	125$\times$122, 61  \\
        						    &   42.4	            &	24.0	            &	15.0 	                &	8.5                &   5.4 & 10.2              \\    
            \hline
        \end{tabular}
        \caption{Image properties for the sample of clusters
            with known diffuse emission studied here.
            Note that the TGSS beam is circular at both resolutions.
        }
        \label{tab:beamRMS}
    \end{table*}
 
    In Table~\ref{tab:beamRMS} we give the RMS values of the GLEAM images
    used in this paper
    as well as those of the TGSS 150 MHz images at two separate resolutions,
    the original $25\arcsec$ images and images tapered to $60\arcsec$
    to highlight the diffuse nature of the emission.
    This table also gives the resolutions of images
    at all the four GLEAM frequencies.
    Due to the poor resolution of the GLEAM images, occasionally unrelated sources
    get blended with the haloes and relics of interest.
    In order to estimate the flux densities of these haloes and relics at the GLEAM
    frequencies the flux densities of the unrelated sources were subtracted.
    Table~\ref{tab:ur} shows the positions of these unrelated sources as well as their
    flux densities at 200 MHz and their spectral indices.
    The integrated flux densities of the haloes and relics 
    are given in Table~\ref{table:flux}.
    Note here that while the TGSS measurements use the calibration scale
    provided by \cite{scaife12} and the MWA uses the calibration scheme of \cite{baars77}
    the difference between the two scales is $\sim3$\% \citep{nhw17}.
    
    We also estimated the angular sizes of the haloes and relics at 200 MHz
    using the task {\tt imfit} in the Common Astronomy Software Analysis (CASA) package
    (Table~\ref{tab:angsize}).
    Also shown in Table~\ref{tab:angsize} are the positions and linear extents
    of the haloes and relics as well as the redshifts of the clusters.

    In Fig.~\ref{fig:cx}, we show the contours of the GLEAM 200 MHz wideband images
    overlaid on the respective X-ray images of the clusters.
    All the X-ray images were obtained from the HEASARC webpage
    \footnote{\url{http://heasarc.gsfc.nasa.gov/xamin/xamin.jsp}}
    or the XMM database \footnote{\url{http://nxsa.esac.esa.int/nxsa-web}}.
    We used {\it XMM-Newton} and {\it Chandra} images where available.
    Although we make use of the images at other MWA frequencies 
    we are only showing the 200 MHz images
    as they have the best resolution and sensitivity.
    Fig.~\ref{fig:cg} shows the GLEAM 200 MHz image contours overlaid
    on the corresponding greyscale TGSS images at $60\arcsec$ resolution.
    The exception to this is A13 for which we have used the $25\arcsec$
    resolution image.
    Fig.\ref{fig:spectra} shows the spectra of all the halos and relics
    that were detected.
    
    The results on the individual clusters are discussed below.

    \subsection{Abell 13}

            \begin{table}
            \centering
            \begin{tabular}{lcccc}
                \hline
                Cluster & \multicolumn{2}{c}{Position}      & $S_{200}$ & $\alpha$  \\
                        &   RA      &   DEC                 & (mJy)     &           \\
                \hline
                A13     & 00:13:33  & -19:28:52             & 101.24    & -2.0      \\
                \hline
                A548b   & 05:45:21  & -25:55:55             & 28.8      & -0.74     \\
                        & 05:45:27  & -25:55:10             & 4.5       & 0         \\
                        & 05:45:11  & -25:54:55             & 22.2      & -0.52     \\
                        & 05:45:22  & -25:47:30             & 74.8      & -0.13     \\
                \hline
                A2163   & 16:16:03  & -06:09:28             & 81.65     & -1.59     \\
                        & 16:15:40  & -06:13:48             & 373.48    & -1.26     \\
                        & 16:15:27  & -06:07:02             & 224.35    & -1.41     \\
                        & 16:15:41  & -06:09:08             & 28.46     & -0.8      \\
                        & 16:16:23  & -06:06:46             & 237.2     & -0.65     \\
                \hline
                A2345   & 21:27:34  & -12:10:58             & 143.25    & -0.8      \\
                        & 21:26:45  & -12:07:29             & 199.22    & -0.8      \\
                \hline
     PLCK G287.0+32.9   & 11:50:43  & -28:00:29             & 20.5      & -0.6      \\
                        & 11:50:40  & -28:01:00             & 14.8      & -0.38     \\
                        & 11:50:33  & -27:58:58             & 2.86      & -0.28     \\
                        & 11:50:34  & -28:00:05             & 6.85      & -0.94     \\
                        & 11:50:50  & -28:02:22             & 12.22     & -0.6      \\
                        & 11:50:56  & -28:01:54             & 20.59     & -0.62     \\
                        & 11:51:00  & -28:04:09             & 51.72     & -0.92     \\
                        & 11:50:52  & -28:05:24             & 37.01     & -0.56     \\
                        & 11:50:46  & -28:05:42             & 147.07    & -0.9      \\
                        & 11:50:59  & -28:00:40             & 34.8      & -1.37     \\
                        & 11:50:50  & -27:59:10             & 56.3      & -0.6      \\
                        & 11:50:54  & -27:59:10             & 72.2      & -0.81     \\
                        & 11:50:59  & -27:59:23             & 21.9      & -1.44     \\
                \hline
            \end{tabular}
            \caption{List of unrelated sources whose flux densities were subtracted from
                the corresponding haloes and/or relics.
                The flux density of the unrelated source at any given frequency is
                $S_{\nu} = S_{200}*(\nu_{\rm MHz}/200)^{\alpha}$.
                The spectral indices were estimated based on the TGSS and the higher
                frequency observations.
                Detailed references for the higher frequency observations are given in
                Table~\ref{table:flux}.
            }
            \label{tab:ur}
        \end{table}

        A low redshift cluster ($z=0.0946$,
        \citealt{strood99}) with a highly disturbed morphology, 
        the X-ray distribution of Abell 13 (A13) shows
        two distinct clumps centred around the two brightest cluster galaxies
        \citep{juett08}.
        The X-ray luminosity of the cluster is
        $L_{\rm X [0.1-2.4 keV]} = 1.24\times10^{44}$ ergs/s \citep{piffaretti11}.
        A13 was observed at radio frequencies by \cite{slee01}
        where they detected an irregularly shaped relic at 1.4 GHz.
        
        The temperature map of the cluster \citep{juett08} shows
        a drop in the temperature at the site of the radio relic.
        This seems to suggest that
        the origin of the relic in A13 is not shock related 
        as the temperature in regions around shock accelerated relics
        is usually greater than in the cluster centres.

        Fig.~\ref{a13_x} shows
        the GLEAM 200 MHz contours of A13 overlaid on
        the corresponding {\it XMM-Newton} image.
        The irregularly shaped object (A) towards the West of the X-ray emission
        is the radio relic in A13.
        The other two sources --
        one to the N and the other to the SE of the X-ray emission --
        are galaxies with optical counterparts and
        are unrelated to the relic emission.
        When compared with the 1.4 GHz image of \cite{slee01},
        the relic is more extended in the GLEAM images.
        This extension is over and above what can be expected
        due to convolution effects.
        The apparently smaller extent of the relic emission at 1.4 GHz 
        is due to the extended emission being resolved out
        in the VLA-BnA configuration used by \cite{slee01}.

        In Fig.~\ref{a13_g} we show the GLEAM 200 MHz contours overlaid
        on the corresponding $25\arcsec$ TGSS greyscale image where
        a faint source embedded within the MWA contours
        is seen to the NE of the relic emission.
        This source is unrelated to the relic and has an optical 
        counterpart.
        This source is not detected in the NVSS 1400 MHz image.
        We estimate the spectral index of this source to be $\sim-2$
        based on the TGSS 150 MHz measurement and the NVSS 1400 MHz upper limit.
        This spectral index is consistent with the flux density estimate of this source
        at 74 MHz from the VLSSr images.
        The extrapolated flux density of this source was subtracted from
        the integrated flux density of source A to estimate the flux densities
        of the relic at all the GLEAM survey frequencies used in this study.
        
        The spectrum of the A13 relic is shown in Fig.~\ref{a13_sr}.
        No diffuse radio emission (halo) associated with the cluster
        X-ray  emission was detected.
        \cite{slee01} produced a spectrum for the relic over the frequency range
        $80-1400$ MHz which showed a curvature that is not seen by our measurements.
        However, it should be mentioned that the curved spectrum of the A13 relic
        as seen by \cite{slee01} is largely influenced by the Culgoora flux density
        measurements at 80 MHz.
        Given the uncertainty in the Culgoora measurements
        we do not believe that the A13 relic has a curved spectrum.
        With the currently available more accurate flux density measurements of the relic
        using MWA we believe the relic spectrum to fit a power law
        and not the curved spectrum as seen by \cite{slee01}.


    \subsection{Abell 548b}
        Abell 548b (A548b) is a dynamically unrelaxed cluster
        at a redshift of $z = 0.0424$ \citep{dengert96} 
        but is not extremely X-ray luminous
        ($L_{\rm X [0.1-2.4 keV]} = 0.1\times10^{44}$ ergs/s, \citealt{piffaretti11}).
    
        The cluster is claimed to contain two radio relics:
        one to the north and the other to the north-west of the cluster centre.
        The NW relic was first detected by \cite{giovannini99} and later confirmed
        through more detailed observations by \cite{feretti06}.
        These two claimed relics are $\sim$ 500 kpc from the cluster centre and
        near the boundary of the X-ray emission
        \citep{solovyeva08}.

        In addition to the above two relics,
        \cite{feretti06} also detected diffuse emission from an irregular source
        closer to the centre of the cluster.
        While not entirely certain of its nature,
        they posit that this source could be the result of
        an internal shock in the cluster 
        whereas the other two relics are believed
        to trace the outgoing shocks generated at the time of cluster merger.
   
        Fig.~\ref{a548b_x} shows the GLEAM 200 MHz contours overlaid on
        the XMM X-ray image of this cluster.
        There are 4 sources of interest
        which we have labelled A, B, C and D in the figure.
        While the source D is a known radio galaxy with an optical
        counterpart (PGC17721), the nature of the sources A and B, that are claimed
        to be cluster relics (\citep{feretti12}) is still
        a mystery.
        We discuss this in detail in a later section of the paper.
        The central diffuse source (C),
        near the northern edge of the X-ray emission, 
        is a complex source.
        High resolution images of this source \citep{feretti06}
        reveal that source C is in fact a combination of 3 objects --
        the outer 2 of which are compact objects even in the ATCA
        (Australia Telescope Compact Array) 2.5 GHz images.
        The central object, however,
        is diffuse in nature and has no optical source associated with it.
        We estimate the spectral index of this diffuse source between 118 MHz and 1.4 GHz
        to be $-0.7\pm0.1$.
        Based on its radio 
        morphology (\citep{feretti06}) and spectral index, this source appears to be 
        a radio galaxy.
        No diffuse radio emission (halo) associated with the X-ray distribution
        was detected in this cluster.
        
   
        In Fig.~\ref{a548b_g} we have overlaid the GLEAM 200 MHz contours
        on the corresponding $60\arcsec$ TGSS image.

    \subsection{Abell 2063}
        Abell 2063 (A2063) is a low redshift ($z = 0.0343$,
        \citealt{hill93}) galaxy cluster with an X-ray luminosity of
        $L_{\rm X [0.1-2.4 keV]} = 1.14\times10^{44}$ ergs/s \citep{reiprich02}.
    

        The GLEAM 200 MHz contours for A2063
        are overlaid on the {\it Chandra} X-ray emission in Fig.~\ref{a2063_x}.
        In Fig.~\ref{a2063_g} the 200 MHz contours are overlaid on the
        corresponding TGSS image.
        The TGSS radio source, which is unresolved at $25\arcsec$ resolution,
        has an optical counterpart
        with a redshift corresponding to that of the galaxy cluster
        and is most likely a galaxy.
        The radio emission shown in Fig.~\ref{a2063_x} is due to
        a combination of this head-tail radio galaxy
        and another radio galaxy co-located with the X-ray emission.
        The latter source, [OL97]1520+087, has been discussed in \cite{kanov06}.
        No diffuse emission in the form of relics or haloes
        were detected in this cluster.
        The claimed relic in this cluster \citep{feretti12}
        is the source 3C318.1 which is 2.4 Mpc from the cluster centre
        and is a relic radio galaxy.
        This source is discussed in detail by \cite{kg94} and is not
        relevant to the present discussion. 

    \subsection{Abell 2163}
        Abell 2163 (A2163) ($z = 0.203$, \citealt{strood99})
        is one of the hottest and  most X-ray luminous clusters known
        ($L_{\rm X [0.1-2.4 keV]} = 20.16\times10^{44}$ ergs/s,
        \citealt{piffaretti11}).
        The X-ray distribution of the cluster shows two primary subclusters --
        one to the west and a secondary one to the north \citep{bourdin11}.
        The primary subcluster of A2163 (west) shows a bullet-like nature
        suggesting a recent merger in the E-W direction.
        The north subcluster appears to have not had any recent interaction
        with the west subcluster and is well separated from it.

        This cluster contains a giant radio halo ($l \sim 3$ Mpc) in the middle
        and a small radio relic north-east of the cluster centre \citep{feretti01}.

        Figs.~\ref{a2163_x} and \ref{a2163_g} show the  GLEAM 200 MHz contours
        of this cluster overlaid on the corresponding {\it XMM-Newton} X-ray image
        and the TGSS image, respectively.
        The central radio halo (A) and the relic (B)
        are detected in all the GLEAM images (see Table \ref{table:flux}
        for their flux densities).
        These flux densities were estimated after subtracting
        the flux densities of unrelated sources
        (Fig.~\ref{a2163_g}). The flux densities of these unrelated
        sources estimated at 150 MHz (TGSS images) and at 1400 MHz (\citep{feretti01})
        were used to estimate their spectra and their flux
        densities at all the GLEAM survey frequencies given in Table \ref{table:flux}.

    \subsection{Abell 2254}
        The galaxy cluster Abell 2254 (A2254) is a rich,
        X-ray luminous cluster
        ($L_{\rm X [0.1-2.4 keV]} = 4.79\times10^{44}$rg ergs/s,
        \citealt{bohringer00})
        located at a redshift of $z = 0.178$ \citep{crawford95}.

        Optical and X-ray analysis of the cluster \citep{girardi11} suggest that
        A2254 is a highly disturbed cluster with a very complex morphology.
        There are two major clumps of galaxies
        separated by $\sim 0.5 h^{-1}_{70}$ Mpc in the east-west direction.
        While the primary X-ray peak coincides with the BCG
        (Brightest Cluster Galaxy),
        the secondary X-ray peak in the east clump does not,
        suggesting that the cluster is still in the middle of an ongoing merger.

        Radio observations of A2254 \citep{govoni01radio} show
        an irregularly shaped radio halo with clumpy features.
        Fig.~\ref{a2254_x} shows
        the GLEAM 200 MHz contours overlaid on the corresponding XMM X-ray image.
        The halo is seen at all the available GLEAM images.
        
        We show the same 200 MHz contours overlaid on
        the corresponding $60\arcsec$ TGSS image in Fig.~\ref{a2254_g}.
        Note that the RMS value in the 200 MHz image of A2254 is $\sim 37$ \mjyb
        which is $\sim$ 5 times larger when compared with the RMS values in the
        other 200 MHz images.
        This is because of the position of this cluster. 
        At a declination of $\sim+20$d,
        A2254 is at the northern-most declination setting for MWA
        which results in a larger synthesized beam and RMS.

    \subsection{Abell 2345}
        Abell 2345 (A2345) is a highly disturbed,
        X-ray luminous ($L_{\rm X [0.1-2.4 keV]} = 3.9\times10^{44}$ ergs/s,
        \citealt{bohringer04}) galaxy cluster. 
        The redshift of the cluster is found to be $z = 0.1789$
        \citep{boschin10}.
        
        \cite{boschin10} also highlighted the complex substructure in the cluster.
        A2345 contains three distinct subclumps as seen in optical and X-ray.
        These clumps lie in the east, south-east and north-west directions.
        
        Extended diffuse emission from A2345
        was first seen by \cite{giovannini99}
        as part of the NRAO VLA Sky Survey (NVSS) 
        and later confirmed through detailed observations
        at 325 MHz and 1.4 GHz by \cite{bonafede09}. 
        The cluster contains two relics (A \& B)
        on opposite sides of the cluster centre,
        both $\sim$ 1 Mpc from the cluster centre. 
        Each relic is about 1 Mpc in size. 

        In Fig.~\ref{a2345_x},
        we show the GLEAM 200 MHz contours overlaid on the corresponding
        XMM X-ray image.
        The two relics are seen
        to the east (A) and to the west (B) of the central X-ray emission.
        The source above the east relic (C),
        which has a counterpart in the $60\arcsec$ TGSS image
        (Fig.~\ref{a2345_g}),
        is an unresolved source, has an infra-red counterpart in the
        2 Micron All Sky Survey images and is unrelated to the relic emission. 
        The source D is a cD galaxy while the source E
        is another radio galaxy in the cluster \citep{bonafede09}.
        The image obtained after subtracting sources D and E
        does not detect any radio halo that is associated with the cluster.
        Table~\ref{table:flux} contains the flux densities for the relics at
        all the GLEAM frequencies except 88 MHz where they get blended with other
        sources.

        \afterpage{
        \clearpage
        \thispagestyle{empty}
        \begin{landscape}
            \begin{table}
        	\centering
	        \resizebox{\linewidth}{!}{
            \begin{tabular}{c|c|ccccccccc|cc}
	        \hline
		    Cluster	&	Object	    &	\multicolumn{9}{|c}{Integrated Flux Densities (mJy)}	& $L_\text{X[0.1-2.4 keV]}$ & log($P_\text{1.4}$)\\
				    &	&	88	MHz & 118 MHz &	150	MHz & 154 MHz &	200	MHz & 325 MHz &	610	MHz & 1400 MHz  & 3000 MHz  & ($\times 10^{44}$) erg s$^{-1}$	& W Hz$^{-1}$\\
	    	\hline
		    A13	& H & -- & -- & -- & -- & -- & -- & -- & -- & -- & 1.24 & $\lesssim$21.79\\
            & R	& $4445.5\pm355.6$ & $2817.4\pm225.4$ &	$1868.1\pm186.8$ &$1861.9\pm149$ & $1290\pm103$ &	 -- & --	& $35.5\pm1.7^{1}$ & --& -- & 24.08\\ \hline
    		A548b& H & -- & -- & -- & -- & -- & -- & -- & -- & -- & 0.1 & $\lesssim$21.00\\
            & B  & $196.05\pm15.68$ & $114.51\pm9.16$ & $142.12\pm14.21$ & $168.26\pm13.46$ & $116.21\pm9.3$ & -- & -- & $60\pm5^{2}$ & -- & -- & 23.51\\
            & A  & $307.2\pm24.6$ & $298.4\pm23.9$ & $203.9\pm20.4$ &$256\pm20.5$ & $230.1\pm18.4$ & -- & -- & $61\pm5^{2}$ & -- & -- & 23.51\\ \hline
	    	A2063 & H & -- & -- & -- & -- & -- & -- & -- & -- & -- & 1.14 & $\lesssim$21.23\\ \hline
            A2163 & H & $2828.3\pm226.3$ & $1178\pm94.2$ & $563\pm56.3$& $1235.4\pm98.8$ & $791.1\pm63.3$  & $861\pm86.1^{3}$ & -- & $155\pm2^{4}$ & -- & 20.16	& 25.61\\
            & R & $444.4\pm35.6$ & $172.9\pm13.8$  & $89.9\pm9$ & $252.4\pm20.2$ & $132.4\pm10.6$& $82\pm8.2^{3}$ & -- & $18.7\pm0.3^{4}$ & -- & -- & 24.68\\ \hline
            A2254 & H &	$1899.8\pm247$ & $904.9\pm117.6$ & $1086.9\pm141.3$ & $607.8\pm79$ & $289.4\pm37.6$  & -- & -- &$33.7\pm3.37^{5}$ & -- & 4.79 & 24.79\\ \hline
	    	A2345 & H & -- & -- & -- & -- & -- & -- & -- & -- & -- & 3.9 & $\lesssim$22.38\\
                  & R(E) & -- & $623.9\pm49.9$ & $555.6\pm55.6$ & $592.3\pm47.4$ & $419.8\pm33.6$ & $188\pm3^{6}$  & -- & $29\pm0.4^{6}$ & -- & -- & 24.73\\
                  & R(W) & -- & $1164.2\pm93.1$ & $1025.1\pm102.5$ &$932.3\pm74.6$ & $747.4\pm59.8$ & $291\pm4^{6}$ & -- & $30\pm0.5^{6}$ & -- & -- & 24.74\\ \hline
            A2744 & H &	-- & $759.3\pm60.7$ & $414.9\pm41.5$ & $512.2\pm41$ & $372.5\pm30$ & $218\pm21.8$ & -- & $57.1\pm0.57^{5}$ & -- & 11.82 & 25.73\\ 
                  & R & -- & $285.3\pm22.8$ & $202.9\pm20.3$ & $203.6\pm16.3$ & $129.2\pm10.3$ & $98\pm10$ & -- & $18.2\pm0.18^{5}$ & -- & --	& 25.24\\ \hline
        PLCK G287.0+32.9  & H & -- & --  & $314\pm31.4$ & -- & -- & $63\pm6^{7}$ & $26\pm2.6^{7}$ & -- & $2.9\pm0.3^{7}$ & 17.2 & $ 25.52$\\
            & R(SE) & $621.4\pm49.7$ & $438.3\pm35.1$ & $382\pm38.2$ & $280.7\pm22.5$ & $268\pm20.2$ & $114\pm11.4^{7}$ & $50\pm5^{7}$  & $25\pm5^{7}$ & $5.2\pm0.52^{7}$ & -- & 25.72\\
            & R(NW) & $855.3\pm224.3$ & $683.9\pm168.1$ & $620\pm62$ & $433.2\pm93.4$ & $356.8\pm68.7$ & $216\pm21.6^{7}$ & $110\pm11^{7}$ & $15.2\pm1.52^{7}$ & --& -- & 25.51\\ \hline
            RXC J1314.5-2515 & H & -- & -- & -- & -- & -- & $40\pm3^{8}$ & $10.3\pm0.3^{9}$& -- & -- & 9.89 & 24.18\\
            & R(E) & -- & $96.6\pm7.7$ & $104.8\pm10.5$ & $38.1\pm7.7$ & $86\pm6.9$ & $52\pm4^{8}$ & $28\pm1.4^{9}$ & $11.1\pm0.3^{10}$ & -- & -- & 24.71 \\
            & R(W) & -- & $425.3\pm34$ & $519.8\pm52$ & $267.7\pm21.4$  & $263.1\pm21.1$ & $137\pm11^{8}$ & $64.8\pm3.2^{9}$ & $20.2\pm0.5^{10}$ & -- &	 & 24.97\\ \hline 
	        \end{tabular}
	        }
            \caption{Integrated flux densities of the haloes (H) and relics (R)
                from the current study. All the 150 MHz values are from the TGSS
                images. The references for the other flux densities are as
                follows: $^1$\protect\cite{slee01} $^2$\protect\cite{feretti06}
                $^3$\protect\cite{feretti04} $^4$\protect\cite{feretti01} 
                $^5$\protect\cite{govoni01radio} $^6$\protect\cite{bonafede09} 
                $^7$\protect\cite{bonafede14} $^8$\protect\cite{venturi13}
                $^9$\protect\cite{venturi07} $^{10}$\protect\cite{feretti05}.
                The X-ray luminosities were obtained from the MCXC meta-catalogue by
                \protect\cite{piffaretti11}.
                The only exception to this is PLCK G287.0+32.9 for which
                the X-ray luminosity was obtained from \protect\cite{plck11}.
                }
            \label{table:flux}
            \end{table}
        \end{landscape}
        \clearpage
    }

    \subsection{Abell 2744}

        The cluster Abell 2744 (A2744)
        is located at a redshift of $z = 0.308$ \citep{strood99}
        and is extremely luminous in X-ray
        ($L_{\rm X [0.1-2.4 keV]} = 11.82\times10^{44}$ ergs/s, \cite{ebeling10}).

        Optical and X-ray analysis of this cluster
        \citep{kempner04,boschin06,owers11} shows that
        it is currently undergoing a merger in the north-south direction.

        A giant radio halo ($\sim 2.4$ Mpc)
        was first seen in the cluster by \cite{govoni01comp,govoni01radio}.
        In addition to the halo, this cluster also contains
        an elongated radio relic north-east of the cluster centre.

        The radio halo and the relic were detected in the GLEAM survey
        at all the observed frequencies (Table~\ref{table:flux}).
        The flux densities of the halo and relic at 88 MHz are not given
        as they get blended with each other at that frequency.
        Fig.~\ref{a2744_x} shows
        the GLEAM 200 MHz contours overlaid on the XMM X-ray image.
        The contours of the radio halo (A) are aligned with the X-ray emission.
        The radio relic (B) is the source NE of the X-ray and halo emission.
        The source SE of the X-ray distribution has an optical counterpart
        and is unrelated to the halo and relic emission.
        The GLEAM 200 MHz contours
        are overlaid on the corresponding $60\arcsec$ TGSS image in
        Fig.~\ref{a2744_g}.

    \subsection{PLCK G287.0+32.9}
        PLCK G287.0+32.9 is a high redshift ($z = 0.39$),
        highly X-ray luminous
        ($L_{\rm X [0.1-2.4 keV]} = 17.2\times10^{44}$ ergs/s,
        \citealt{plck11}) galaxy cluster 
        which contains radio relics \citep{bagchi11}
        on opposite sides of the cluster centre to the north and south.

        Subsequent observations of the cluster \citep{bonafede14}
        revealed that the source identified by \cite{bagchi11} 
        as the north relic was in fact the lobe emission from a galaxy.
        The true north relic was located south-west of this emission.
        The projected separation between the relics was estimated
        to be $\sim 3.2$ Mpc which is one of the largest known separations
        between double relics.
        The cluster also contains a radio halo near the cluster centre
        \citep{bonafede14}.
        
        In Fig.~\ref{plck_x}
        we show the GLEAM 200 MHz contours overlaid on
        the corresponding {\it XMM-Newton} image.
        The south-east relic (SE), the north-west relic (NW) and the halo (H)
        are marked in the figure.
        Also shown are the galaxy (G) and the region
        further to the north-west (F) which get blended with the north-west
        relic and halo at frequencies below 200 MHz.
        GMRT observations of the cluster by \cite{bonafede14}
        at 150 ($50\arcsec$), 325 ($13\arcsec$) and 610 MHz ($7\arcsec$)
        show the presence of 10 unresolved sources near the halo and
        north-west relic region.
        In order to estimate the flux density of the north-west relic
        we first had to subtract the flux densities of all these unrelated sources.
        The flux densities of these unrelated sources
        at GMRT frequencies are available in \cite{bonafede14}
        and the extrapolated flux densities at MWA frequencies
        were subtracted from the total flux density in this region.
        Subtracting all the unrelated sources
        still leaves the halo and the north-west relic blended.
        We extrapolated the halo flux density to the MWA frequencies
        using the flux density values given in \cite{bonafede14}
        (Table~\ref{table:flux})
        and subtracted it from the remaining flux density.
        This is quoted as the flux density
        of the north-west relic in Table~\ref{table:flux}.

        Fig.~\ref{plck_g} shows the GLEAM 200 MHz contours
        overlaid on the corresponding $60\arcsec$ TGSS 150 MHz image.
        Note that the TGSS image also detects the radio halo.
 
        In Fig.~\ref{plck_sh} and \ref{plck_ss} we plot the spectra for the
        north-west and the south-east relic, respectively.
        
    \subsection{RXC J1314.4-2515}
        A highly X-ray luminous cluster
        ($L_{\rm X [0.1-2.4 keV]} = 9.89\times10^{44}$ ergs/s,
        \citealt{piffaretti11}),
        RXC J1314.4-2515 is at a redshift of $z = 0.247$ \citep{valtchanov02}.
        The X-ray emission is distorted in the NW-SE direction
        and is not centred on the two BCGs in the cluster \citep{valtchanov02}.

        Radio observations of the cluster \citep{feretti05,venturi07}
        reveal the presence of two radio relics -- one to the east and another
        to the west of the cluster
        and a radio halo in the middle with a bridge to the west relic.

\begin{figure*}
        \centering
        \subfloat[A13\label{a13_x}]{\includegraphics[width=0.6\columnwidth]{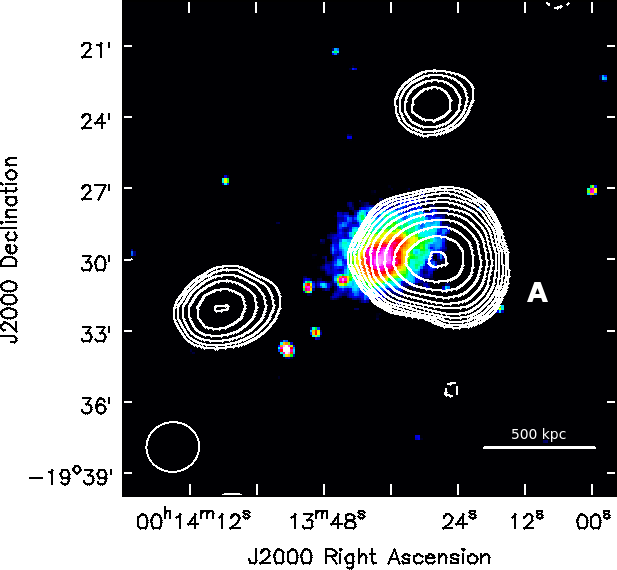}}\hspace{0.5em}
        \subfloat[A548b\label{a548b_x}]{\includegraphics[width=0.6\columnwidth]{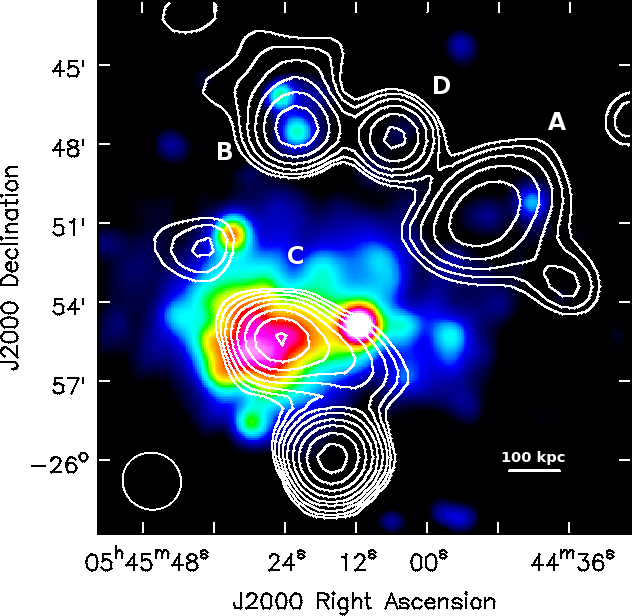}}\hspace{0.5em}
        \subfloat[A2063\label{a2063_x}]{\includegraphics[width=0.6\columnwidth]{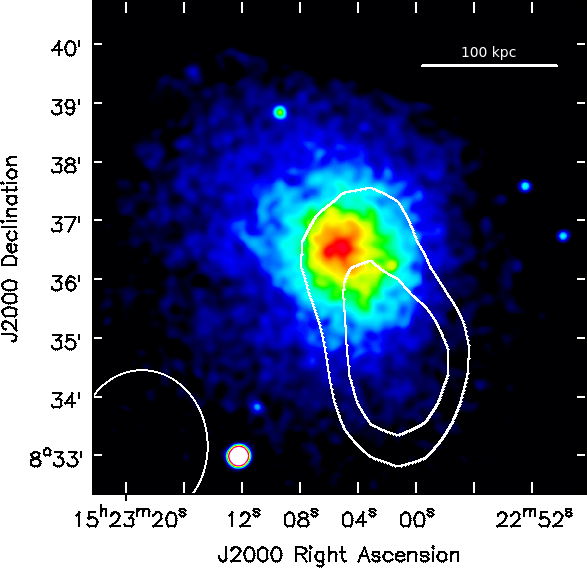}}\\
        \subfloat[A2163\label{a2163_x}]{\includegraphics[width=0.6\columnwidth]{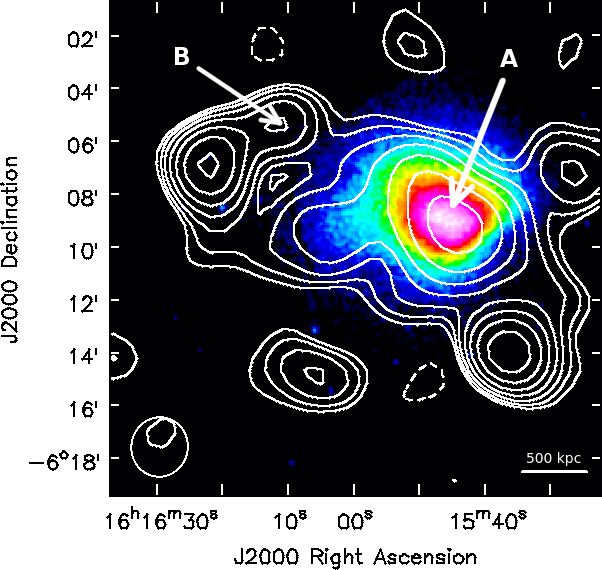}}\hspace{0.5em}
        \subfloat[A2254\label{a2254_x}]{\includegraphics[width=0.6\columnwidth]{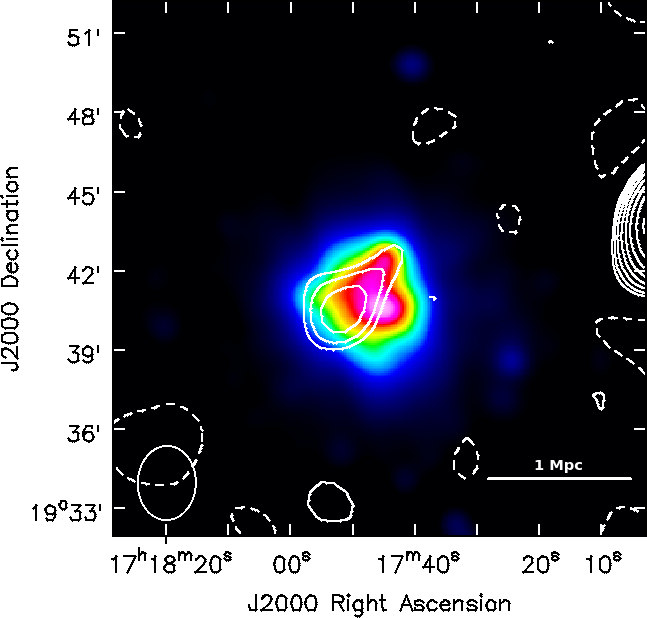}}\hspace{0.5em}
        \subfloat[A2345\label{a2345_x}]{\includegraphics[width=0.6\columnwidth]{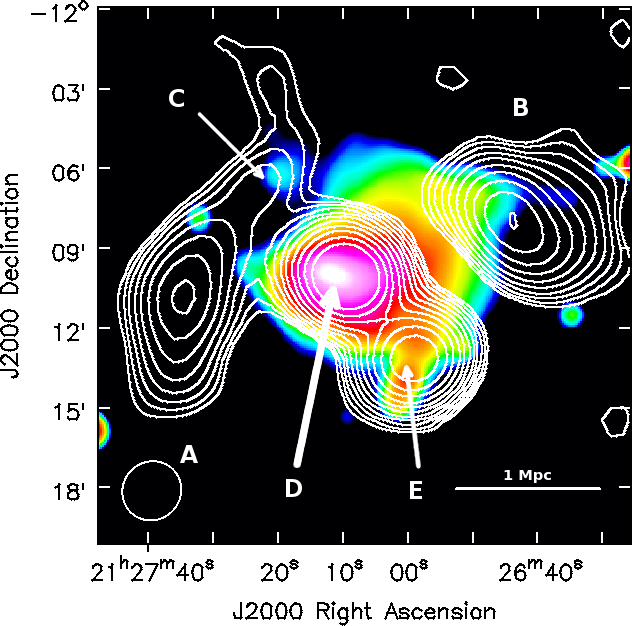}}\\
        \subfloat[A2744\label{a2744_x}]{\includegraphics[width=0.6\columnwidth]{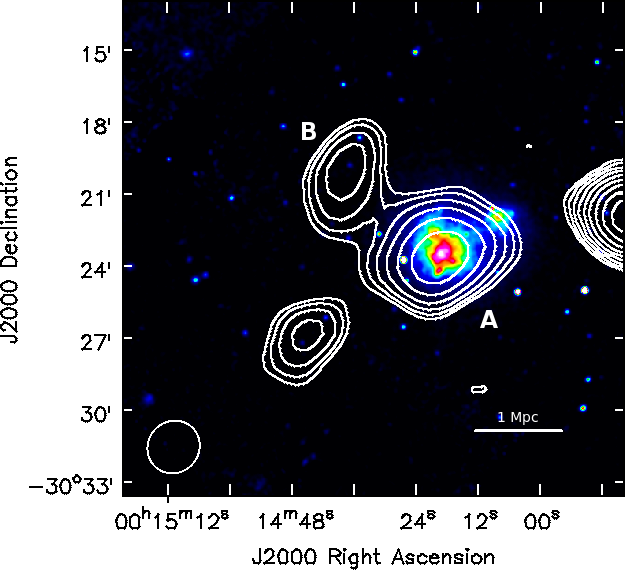}}\hspace{0.5em}
        \subfloat[PLCK G287.0+32.9\label{plck_x}]{\includegraphics[width=0.6\columnwidth]{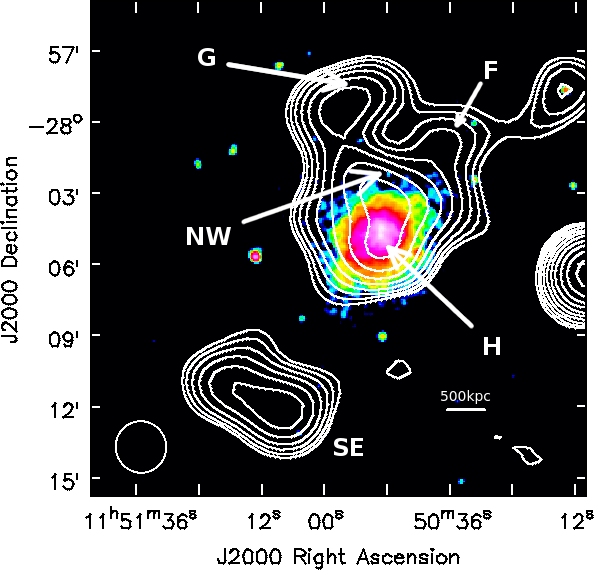}}\hspace{0.5em}
        \subfloat[RXC J1314.4-2515\label{rxcj_x}]{\includegraphics[width=0.6\columnwidth]{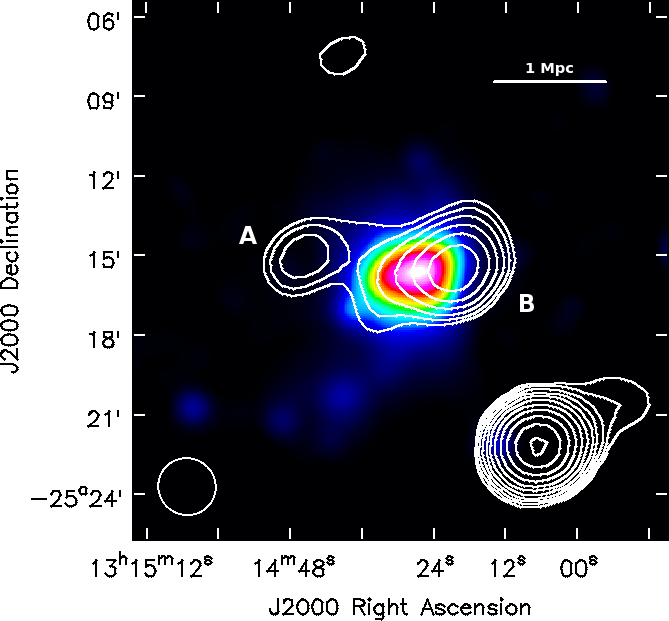}} 
        \caption{GLEAM 200 MHz contours (60 MHz bandwidth) overlaid on
                the respective X-ray images of the clusters. 
                The contours start at $3\sigma$ (at $2\sigma$ for A2254)
                and increase by $\sqrt{2}$ thereafter.
                See Table~\ref{tab:beamRMS} for $\sigma$ values.
                The first negative contour at $3\sigma$ (at $2\sigma$ for A2254)
                is also plotted (dashed lines).
                The full-width half maximum of the synthesized beam of the 
                MWA is indicated in the bottom left-hand corner.
                }
        \label{fig:cx}
    \end{figure*}

        Fig.~\ref{rxcj_x} shows the GLEAM 200 MHz contours of the cluster
        overlaid on the corresponding XMM X-ray image.
        The contours of the east relic (A)
        are seen towards the edge of the X-ray emission.
        The source seen to the west of the X-ray emission (B)
        is a convolution of the west relic and the radio halo in the cluster.
        Even at 200 MHz it is not possible to separate
        the halo emission from that of the west relic.
        Estimating the flux density of the east relic was straightforward.
        In order to estimate the flux density of the west relic
        we first extrapolated the flux density of the halo at 610 MHz
        \citep{venturi07} to the MWA frequencies (up to 118 MHz)
        by assuming a spectral index of $-1.34$.
        This spectral index of $-1.34$ is the average spectral index of haloes from
        literature \citep{feretti12}.
        This extrapolated halo flux density was subtracted
        from the total flux density of the region B
        in order to get the flux density of the west relic.
        At 88 MHz, however,
        the east relic gets blended with the west relic and the halo
        and hence their flux densities are not estimated
        at 88MHz (Table~\ref{table:flux}).

\begin{figure*}
        \centering
        \subfloat[A13\label{a13_g}]{\includegraphics[width=0.6\columnwidth]{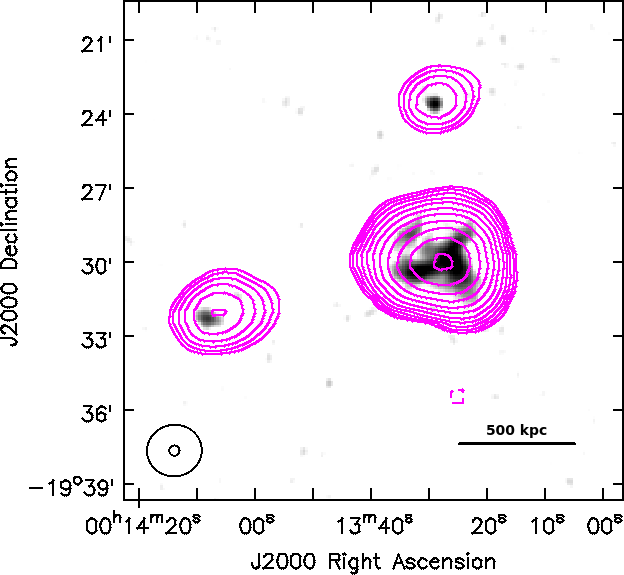}}\hspace{0.5em}
        \subfloat[A548b\label{a548b_g}]{\includegraphics[width=0.6\columnwidth]{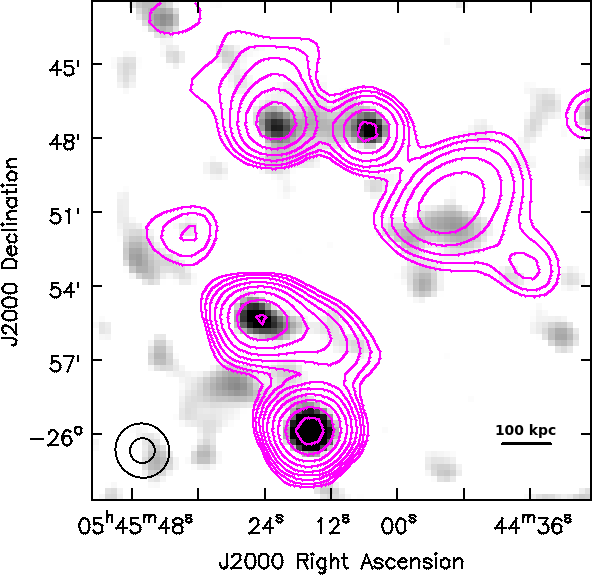}}\hspace{0.5em}
        \subfloat[A2063\label{a2063_g}]{\includegraphics[width=0.6\columnwidth]{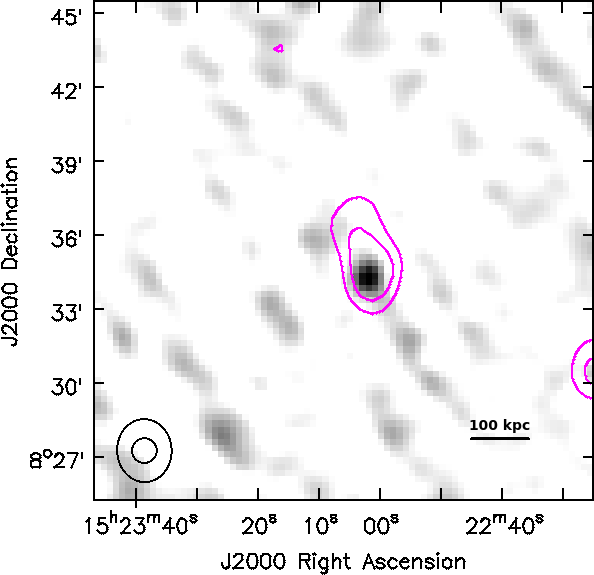}}\\
        \subfloat[A2163\label{a2163_g}]{\includegraphics[width=0.6\columnwidth]{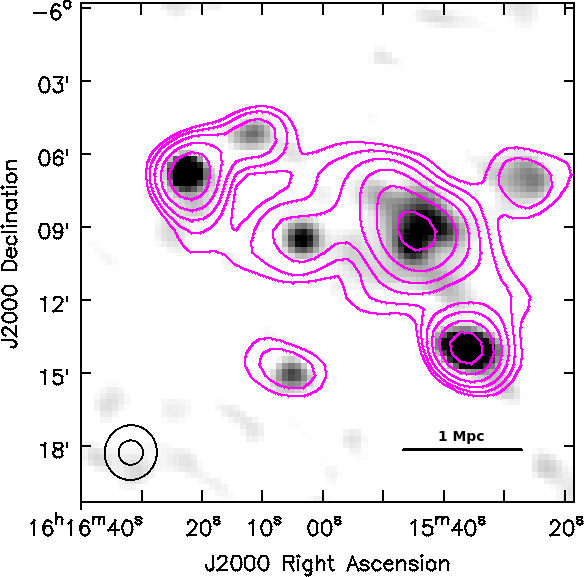}}\hspace{0.5em}
        \subfloat[A2254\label{a2254_g}]{\includegraphics[width=0.6\columnwidth]{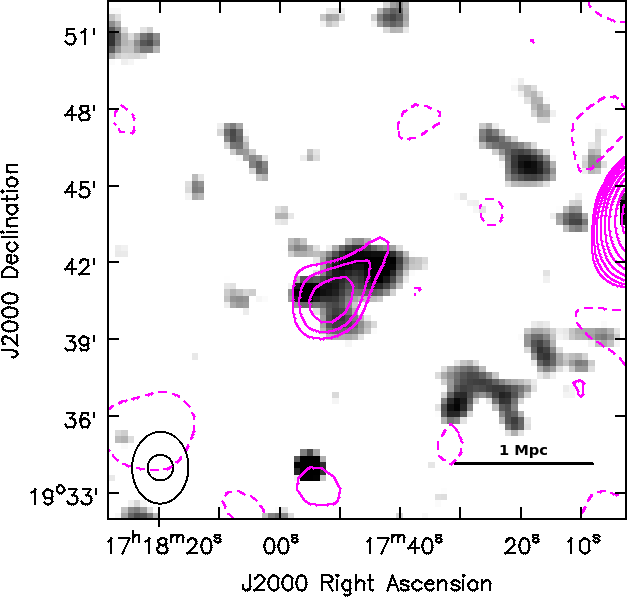}}\hspace{0.5em}
        \subfloat[A2345\label{a2345_g}]{\includegraphics[width=0.6\columnwidth]{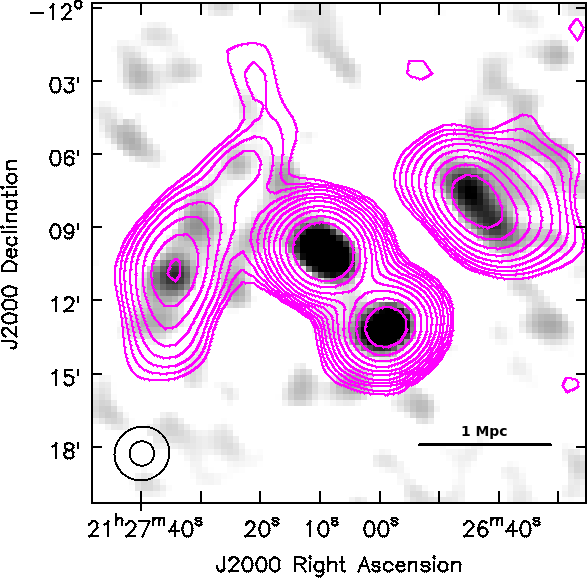}}\\
        \subfloat[A2744\label{a2744_g}]{\includegraphics[width=0.6\columnwidth]{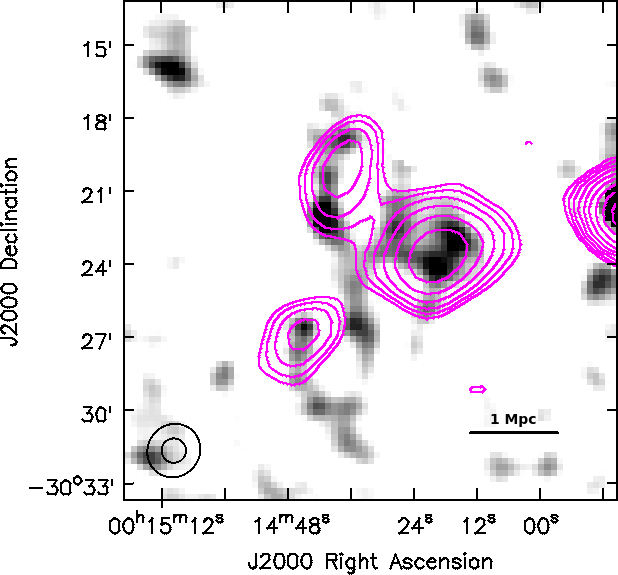}}\hspace{0.5em}
        \subfloat[PLCK G287.0+32.9\label{plck_g}]{\includegraphics[width=0.6\columnwidth]{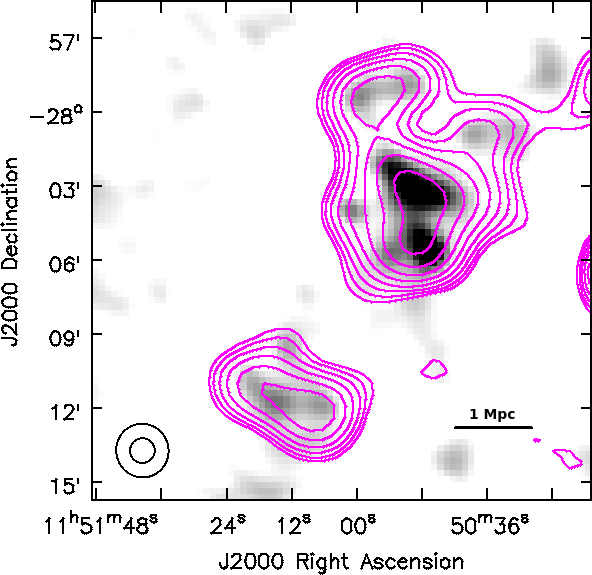}}\hspace{0.5em}
        \subfloat[RXC J1314.4-2515\label{rxcj_g}]{\includegraphics[width=0.6\columnwidth]{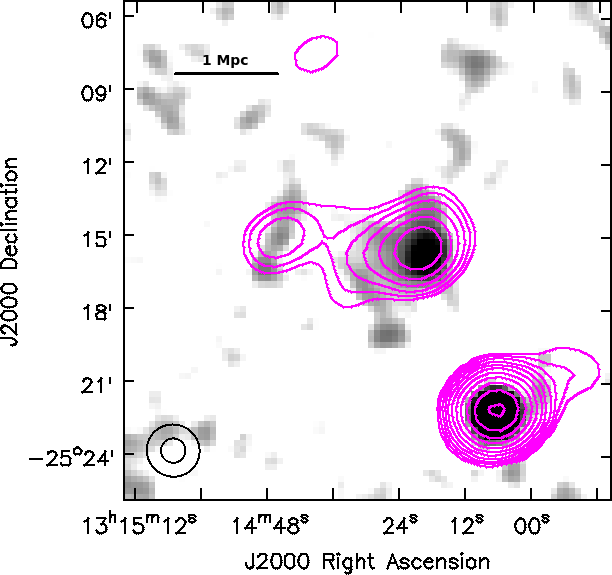}}
        \caption{ GLEAM 200 MHz (60 MHz bandwidth) contours overlaid on
                the corresponding greyscale TGSS 150 MHz images.
                All the TGSS images are at $60\arcsec$ resolution
                except for A13 which is at $25\arcsec$ resolution.
                Contours start at $3\sigma$ (at 2$\sigma$ for A2254)
                and increase by $\sqrt{2}$ thereafter.
                The first negative contour at $3\sigma$ (at $2\sigma$ for A2254)
                is also plotted (dashed lines).
                The full-width half maxima of the synthesized beams of the 
                GLEAM and TGSS images are indicated in the bottom left-hand
                corner.
                }
        \label{fig:cg}
    \end{figure*}

\begin{figure*}
        \centering
        \subfloat[\label{a13_sr}]{\includegraphics[width=0.43\columnwidth,angle=270]{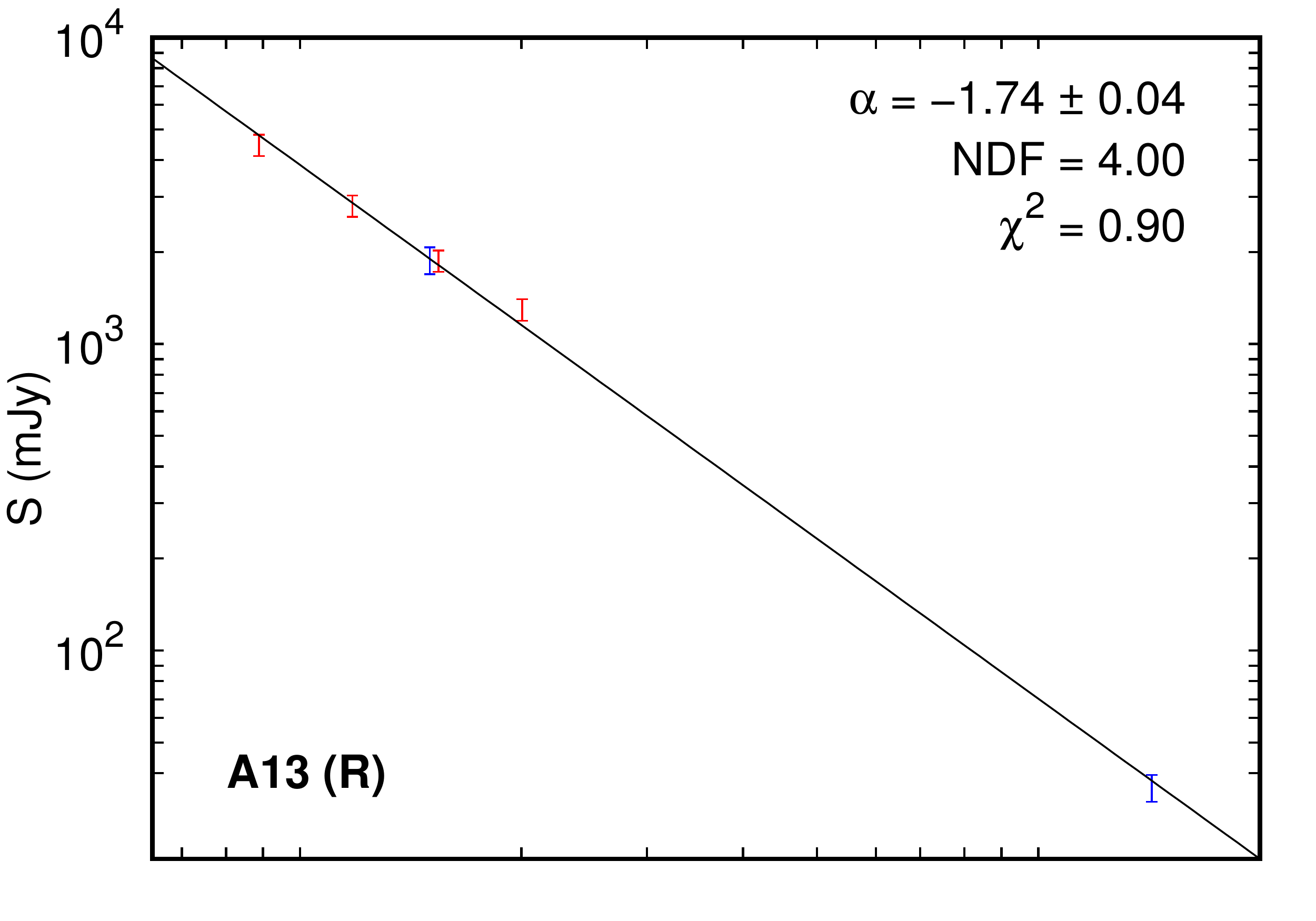}} 
        \subfloat[\label{a548b_sn}]{\includegraphics[width=0.43\columnwidth,angle=270]{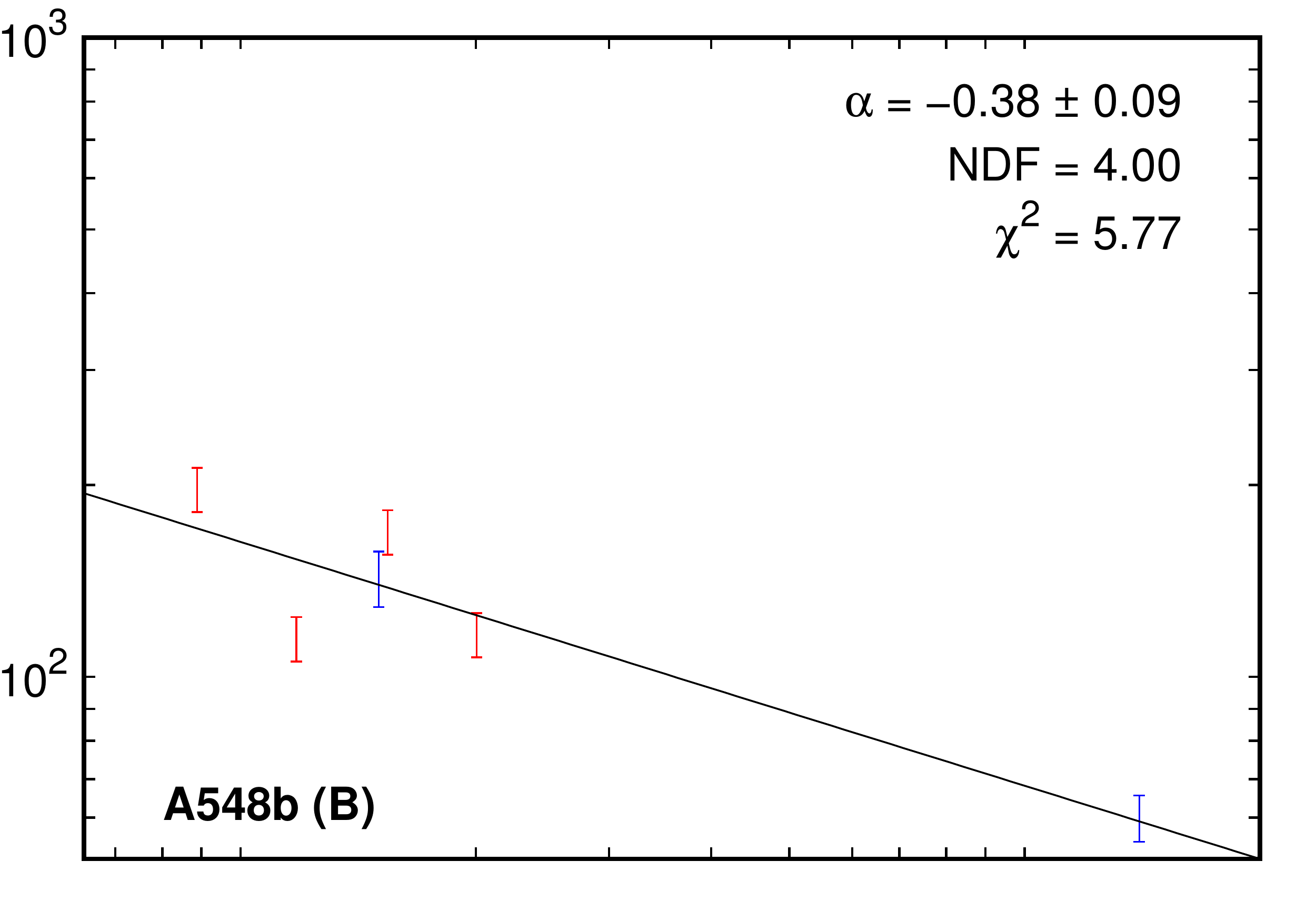}}
        \subfloat[\label{a548b_snw}]{\includegraphics[width=0.43\columnwidth,angle=270]{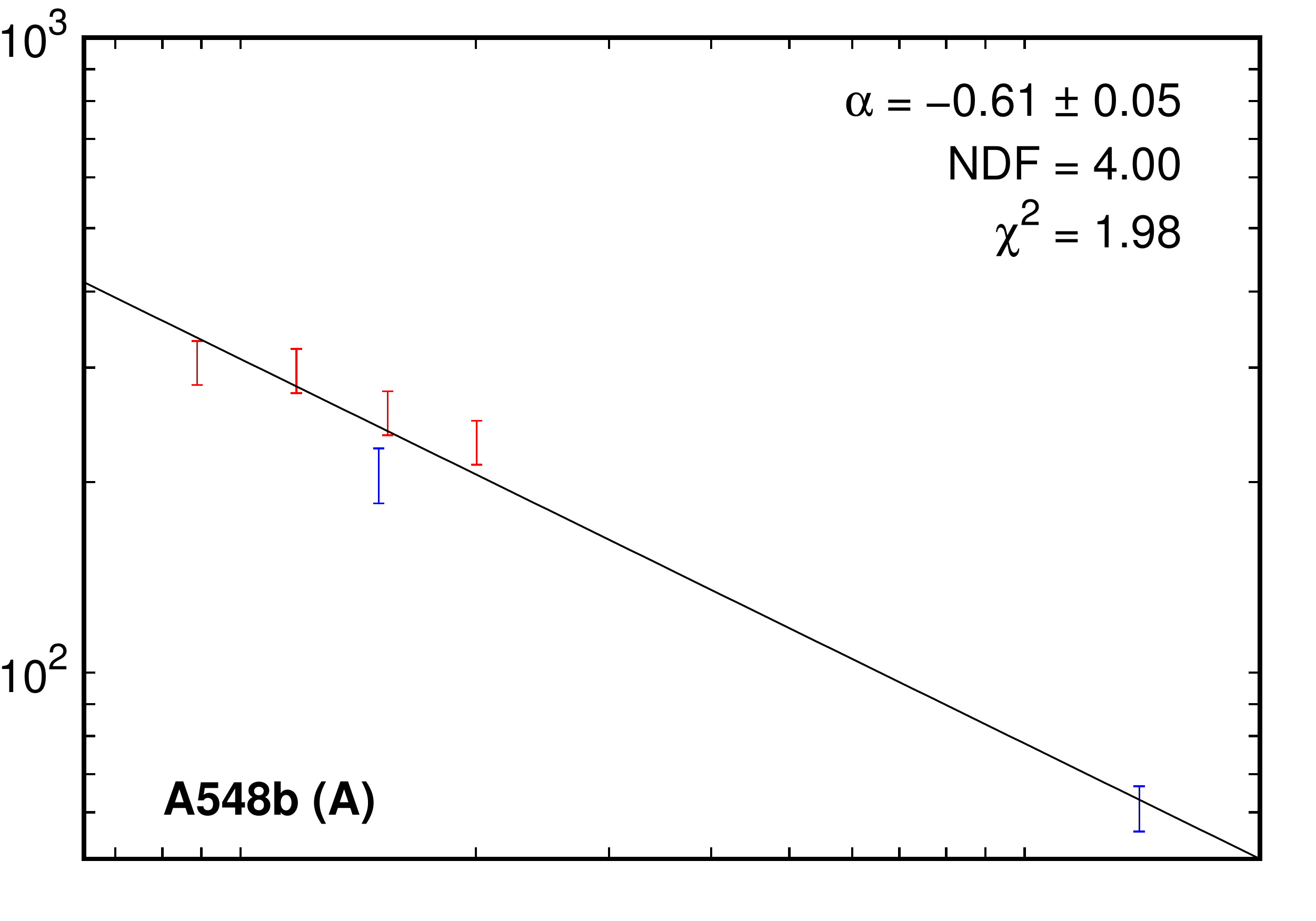}}\\
        \subfloat[\label{a2163_sh}]{\includegraphics[width=0.43\columnwidth,angle=270]{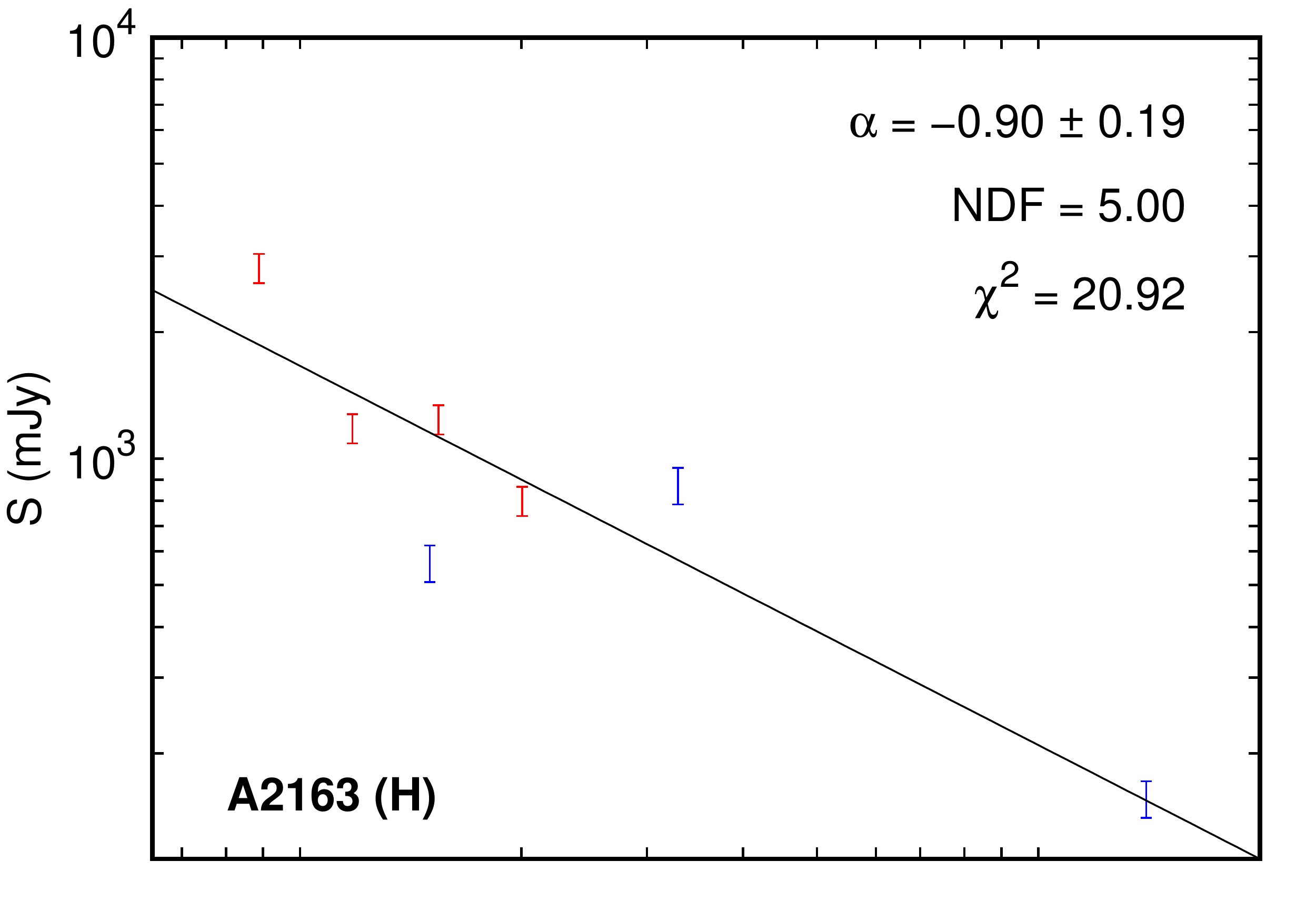}}
        \subfloat[\label{a2163_sr}]{\includegraphics[width=0.43\columnwidth,angle=270]{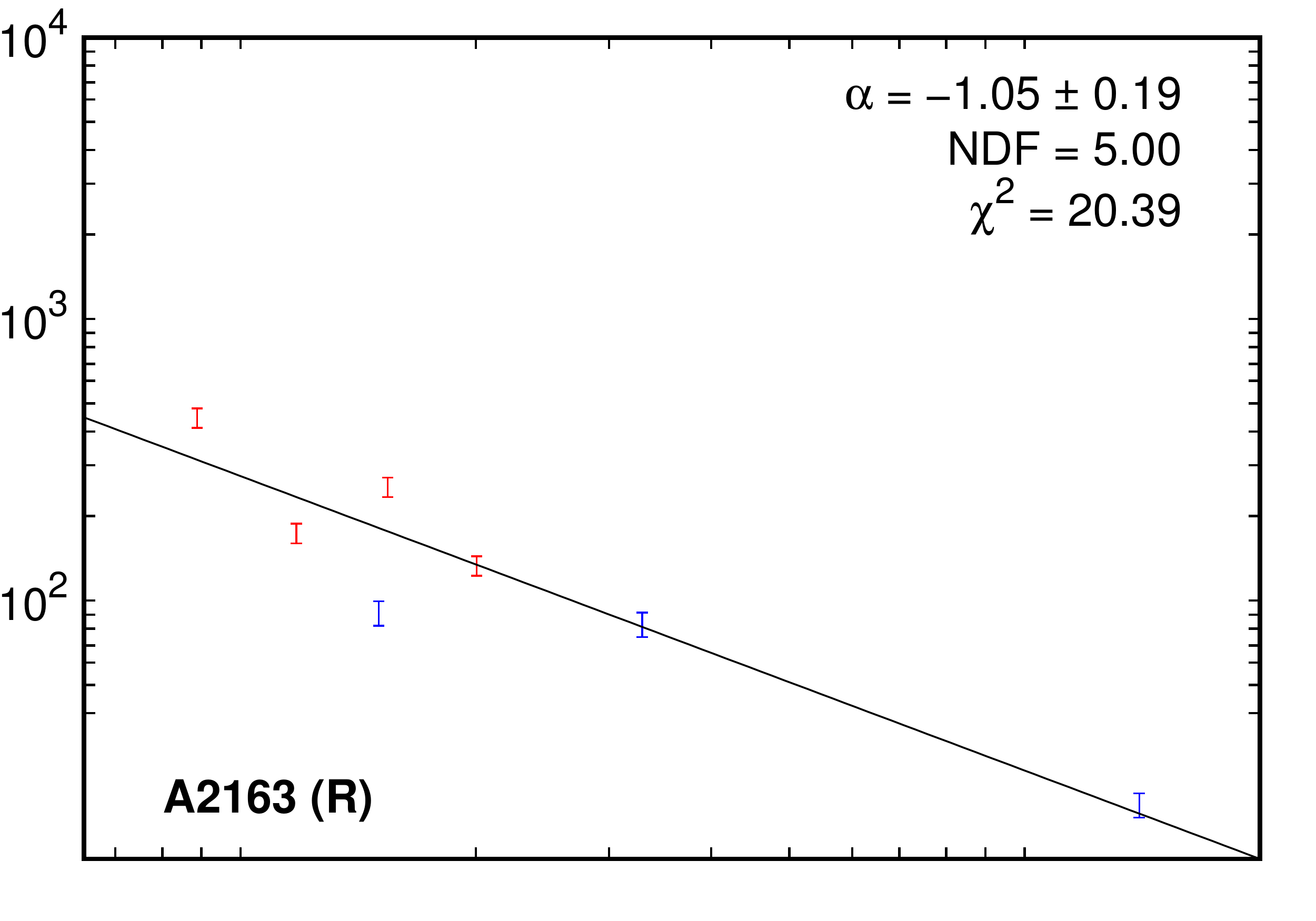}} 
        \subfloat[\label{a2254_sh}]{\includegraphics[width=0.43\columnwidth,angle=270]{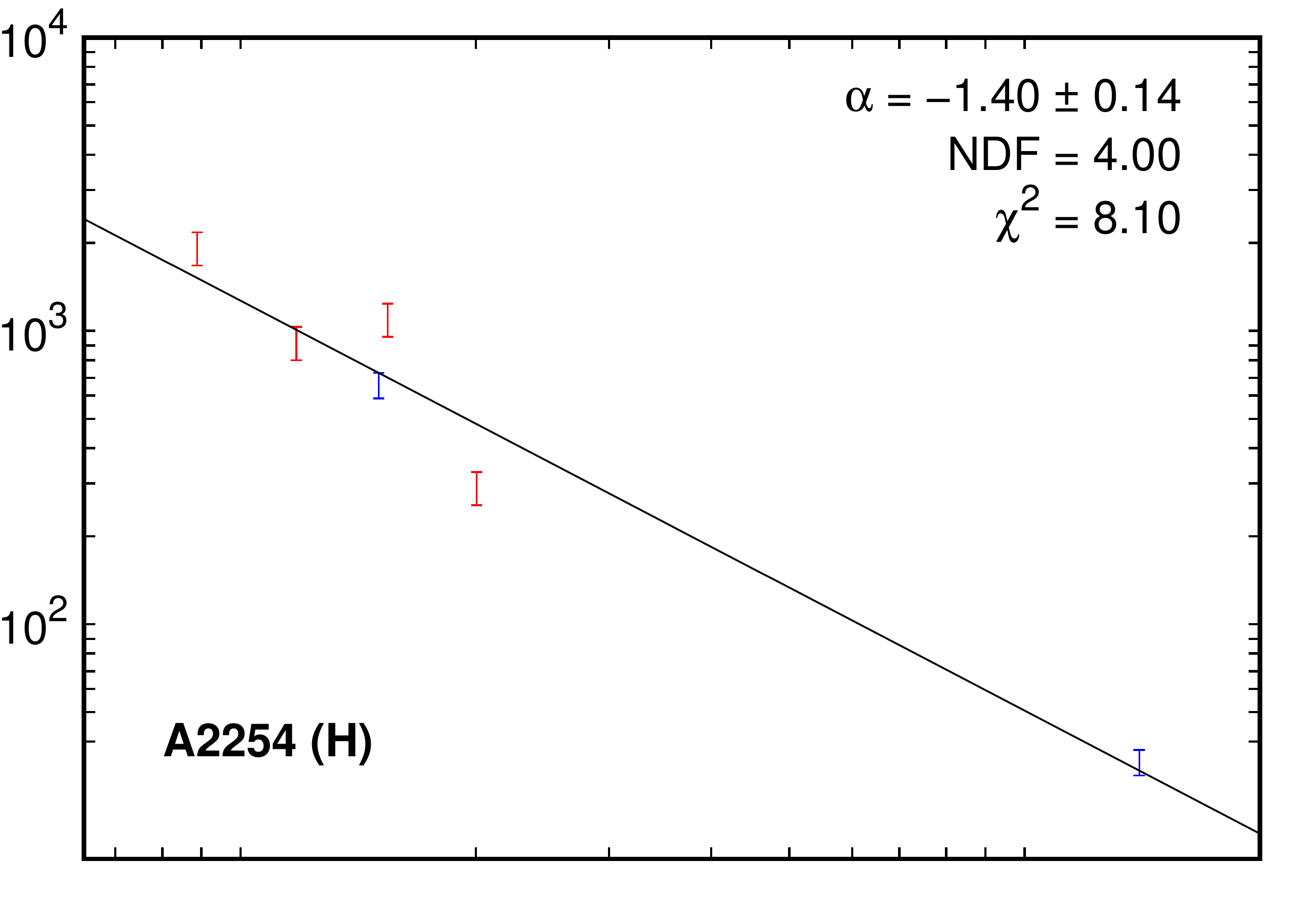}}\\ 
        \subfloat[\label{a2345_se}]{\includegraphics[width=0.43\columnwidth,angle=270]{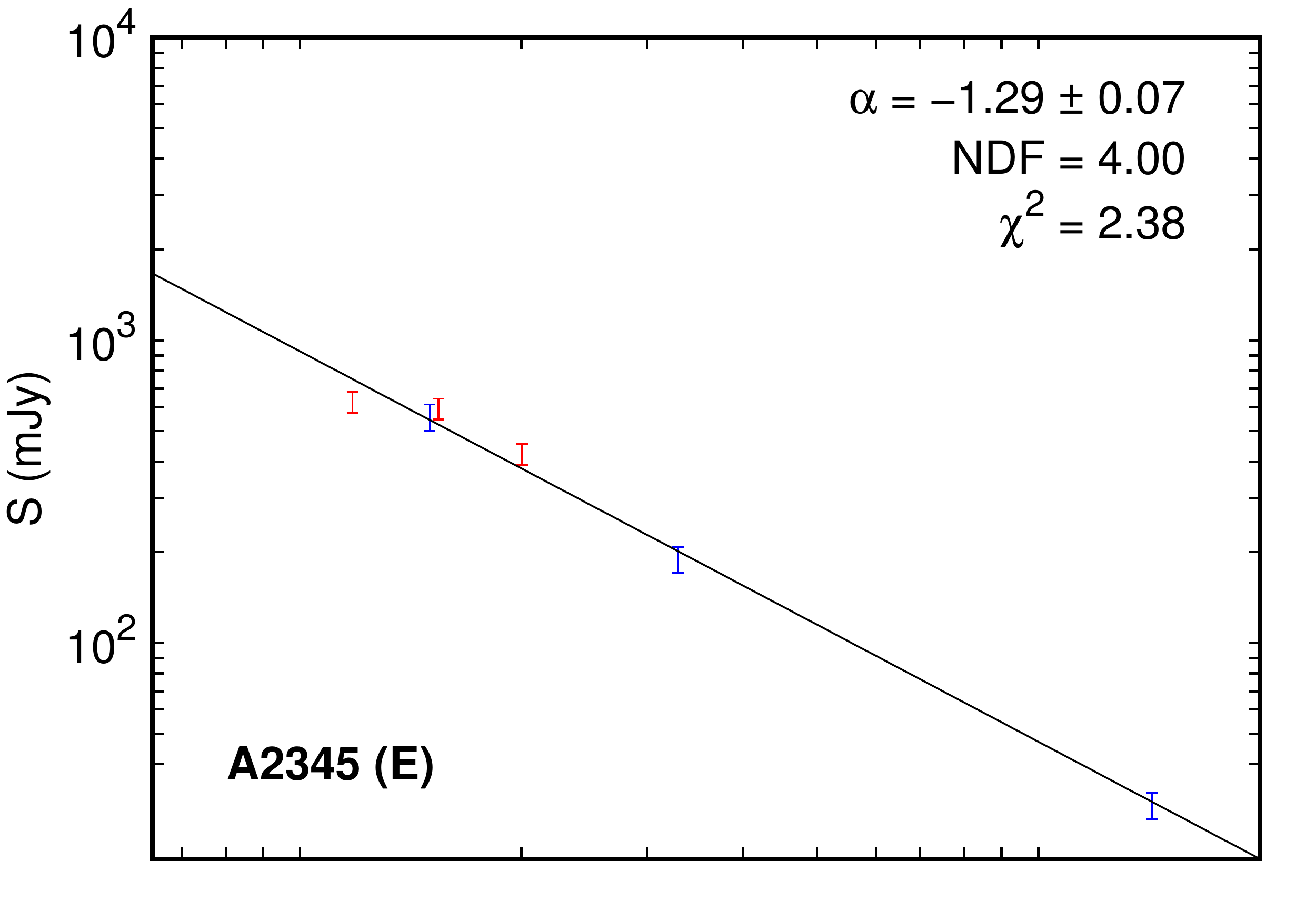}}
        \subfloat[\label{a2345_sw}]{\includegraphics[width=0.43\columnwidth,angle=270]{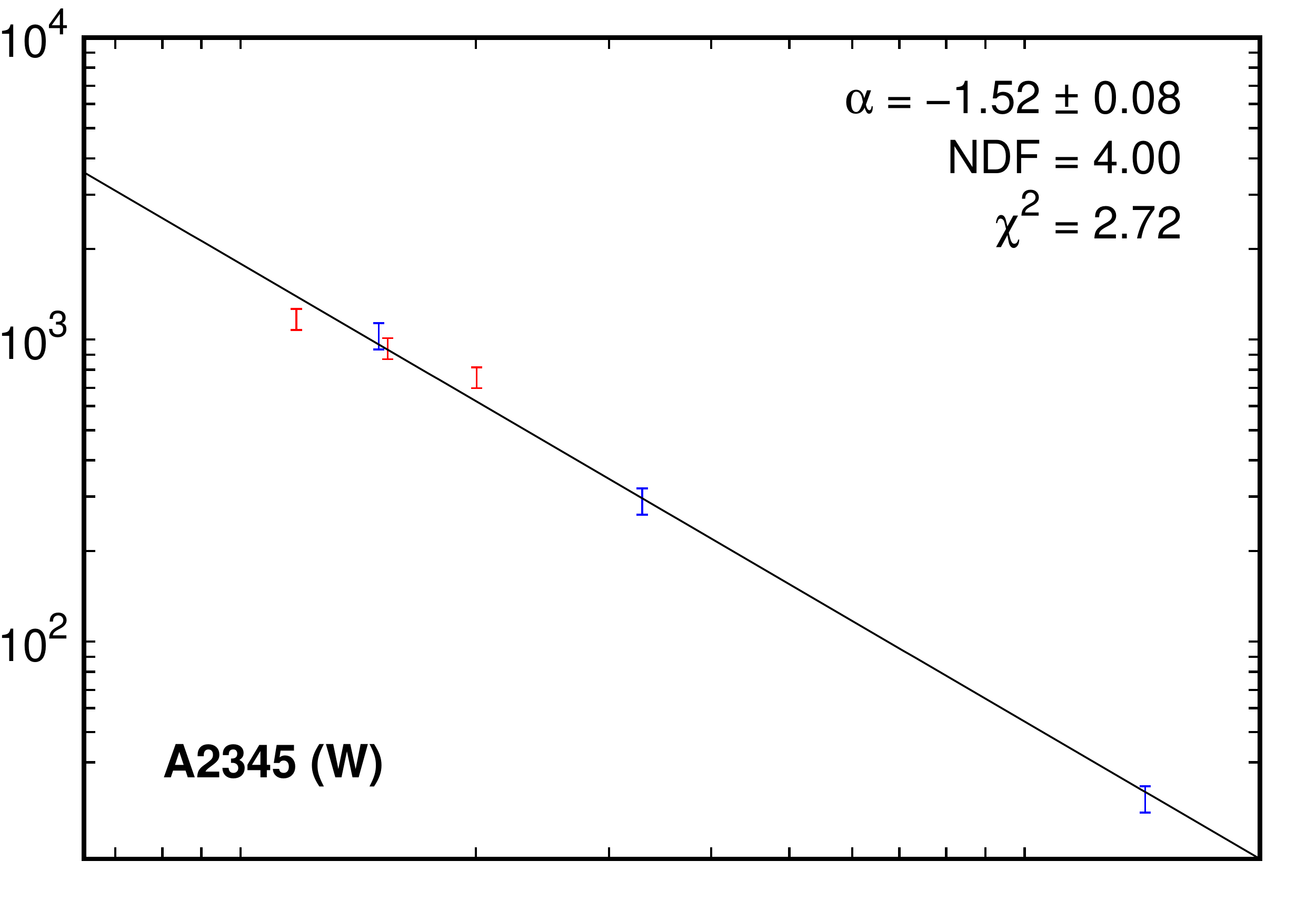}}
        \subfloat[\label{a2744_sh}]{\includegraphics[width=0.43\columnwidth,angle=270]{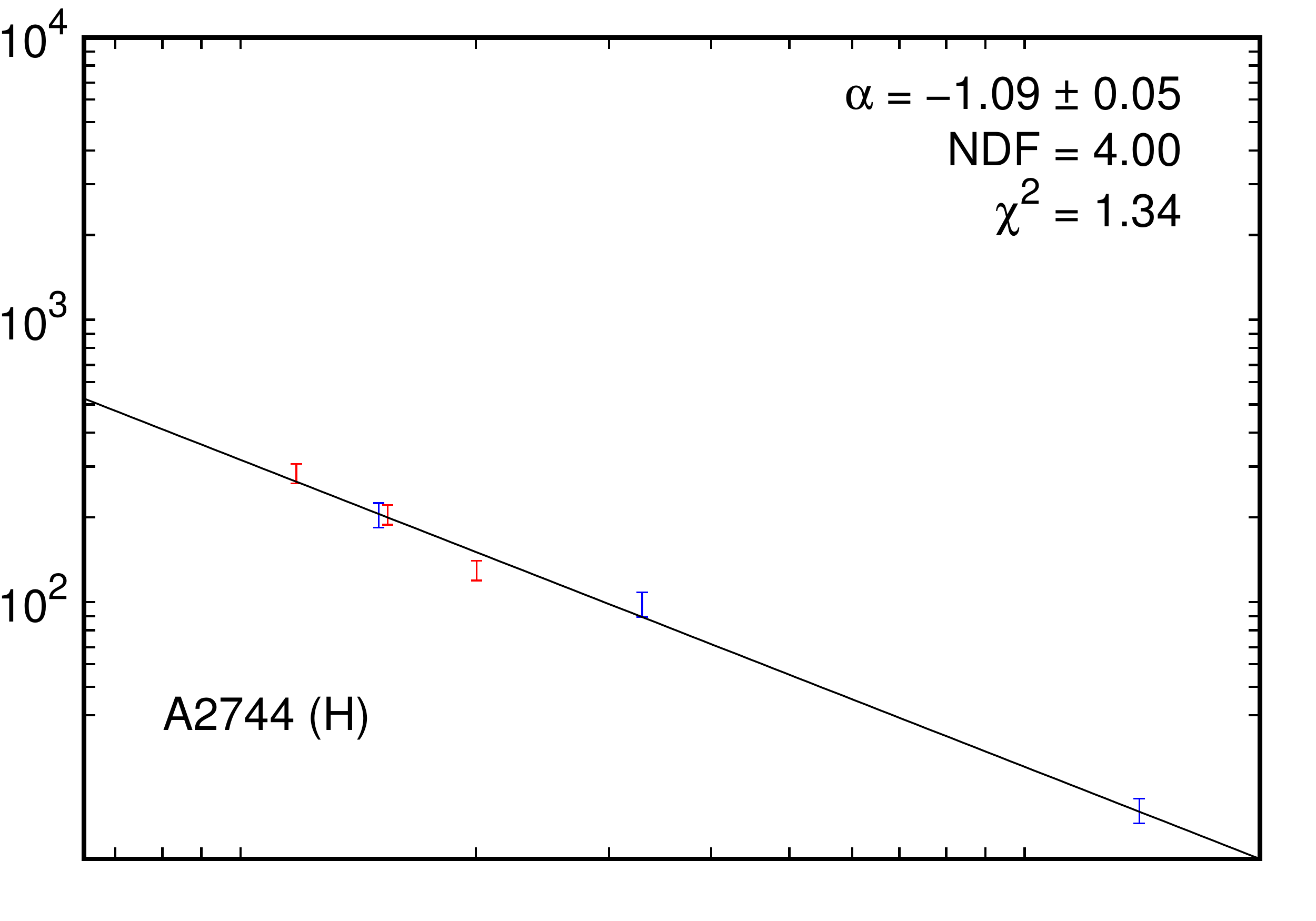}}\\
        \subfloat[\label{a2744_sr}]{\includegraphics[width=0.43\columnwidth,angle=270]{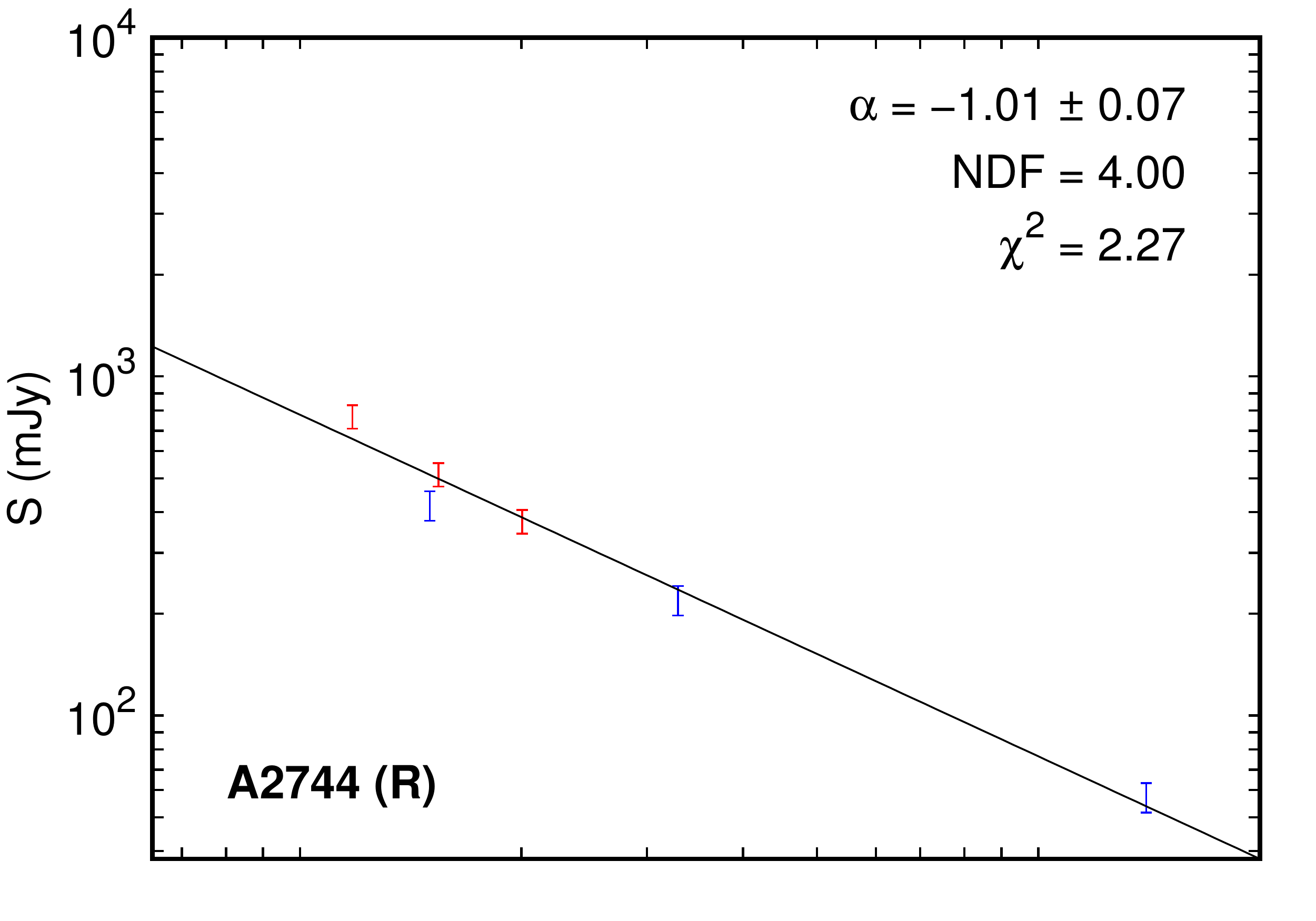}}
        \subfloat[\label{plck_sh}]{\includegraphics[width=0.43\columnwidth,angle=270]{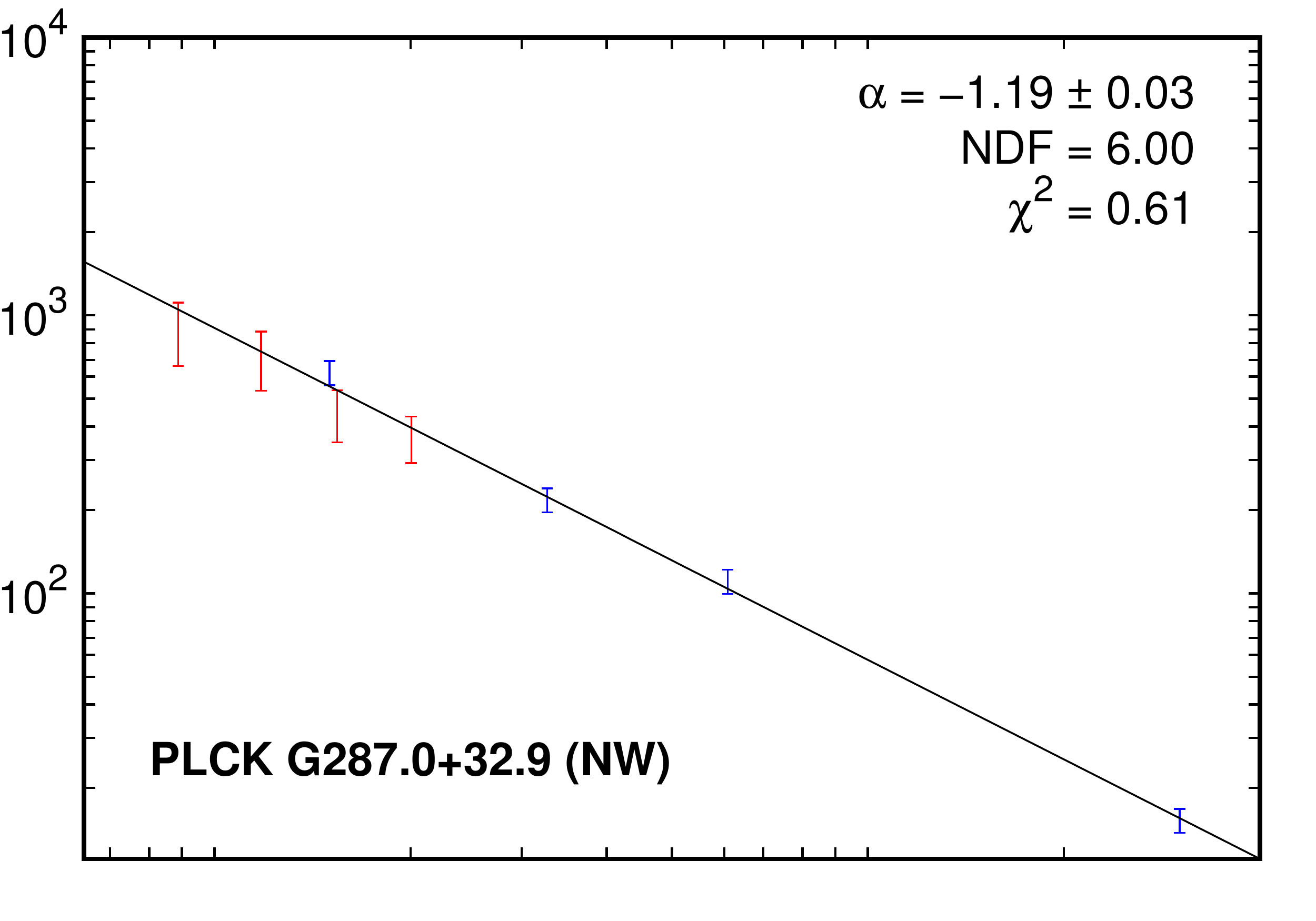}}
        \subfloat[\label{plck_ss}]{\includegraphics[width=0.43\columnwidth,angle=270]{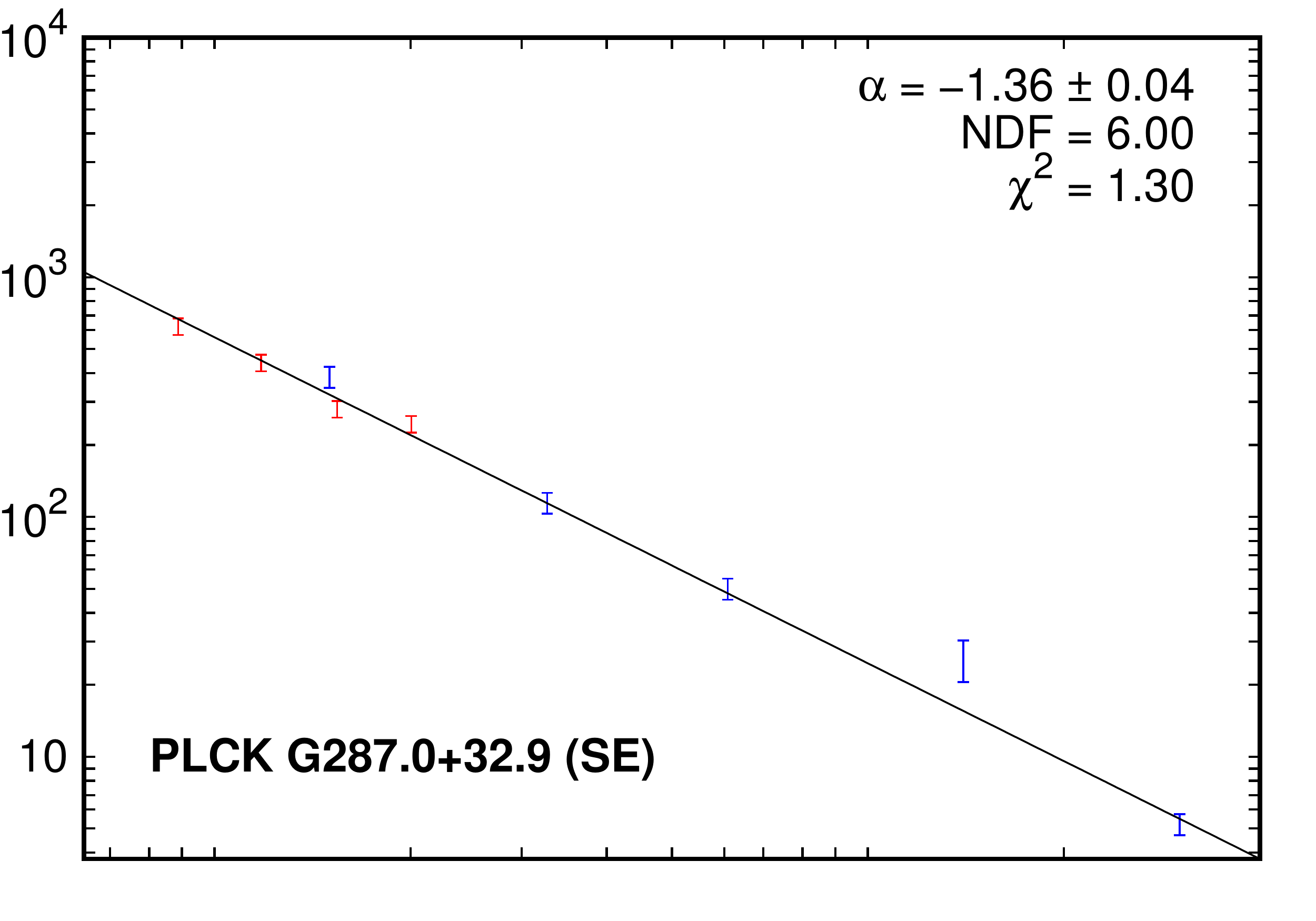}}\\
        \subfloat[\label{rxcj_se}]{\includegraphics[width=0.43\columnwidth,angle=270]{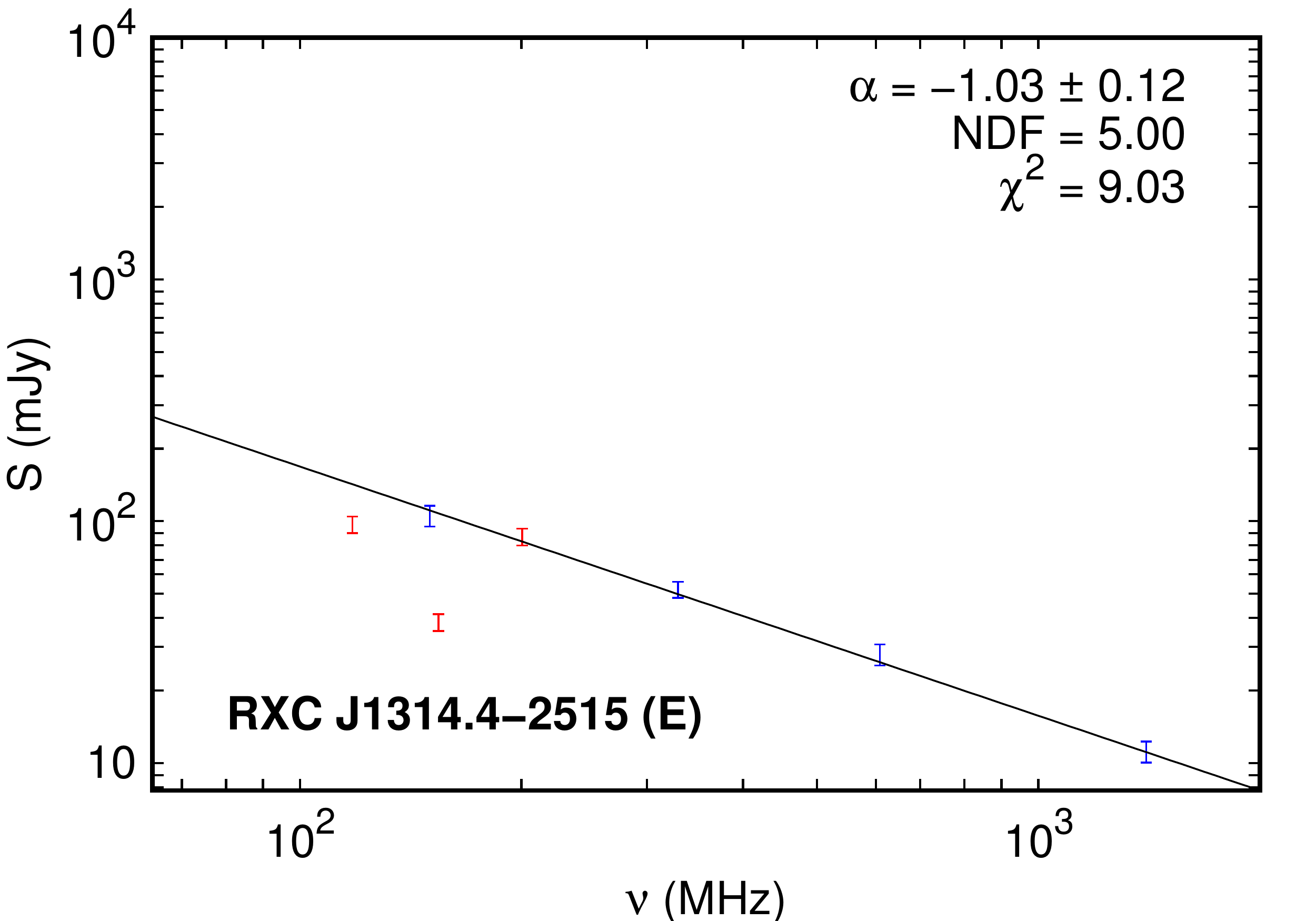}}
        \hspace{5.1cm}
        \subfloat[\label{rxcj_sw}]{\includegraphics[width=0.43\columnwidth,angle=270]{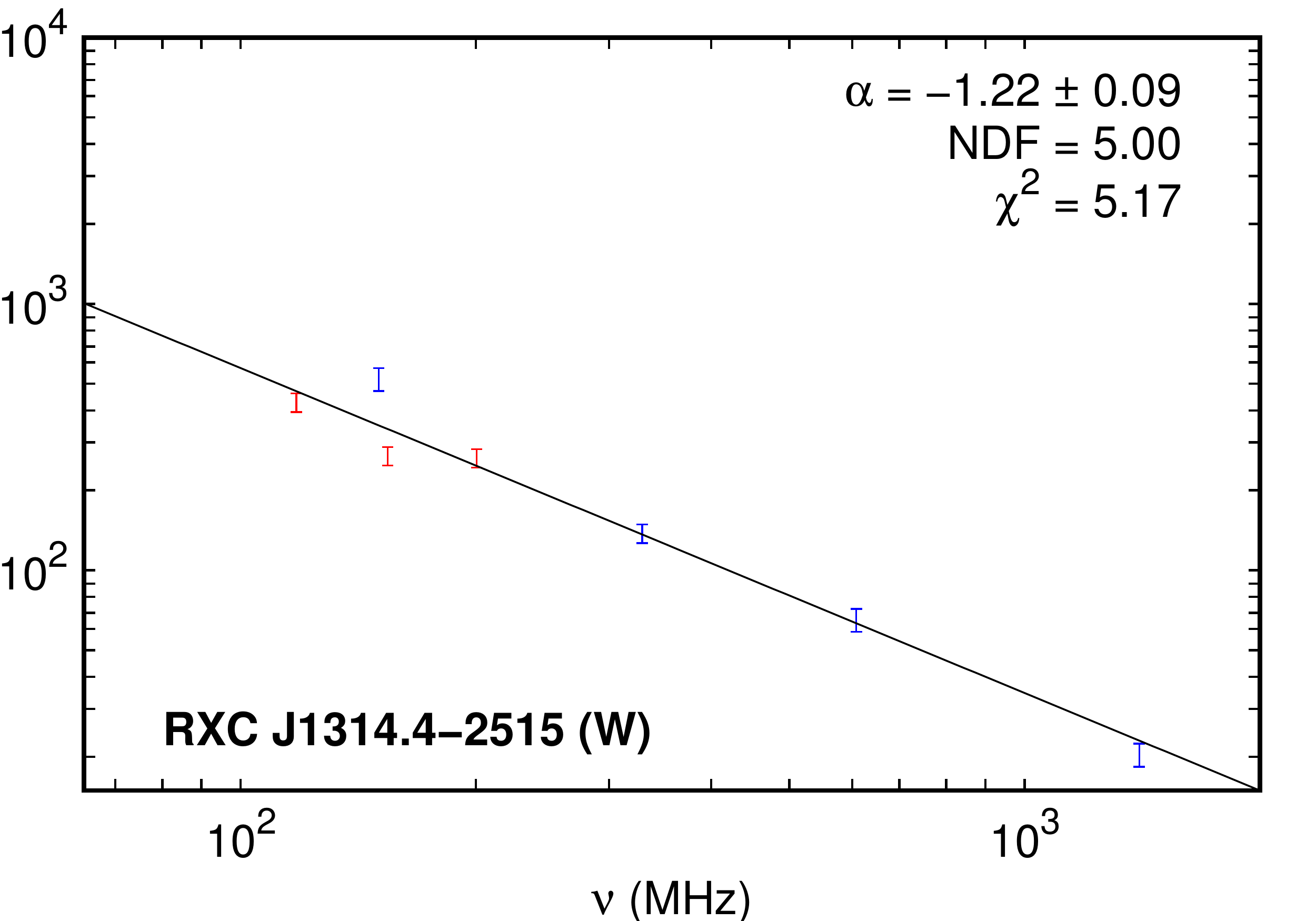}}
        \caption{Spectra of haloes and relics.
        Data points in red colour are from the GLEAM survey
        measurements while those in blue colour are measurements from other telescopes.
        Note that the range of frequencies along the x-axis is identical
        in all the panels and is specified in the bottom-most panel.
        Every panel also shows the spectral index value ($\alpha$),
        the number of degrees of freedom (NDF)
        and the reduced $\chi^{2}$ value of the fit.
    }
        \label{fig:spectra}
    \end{figure*}

        
        In Fig.~\ref{rxcj_g} we show the contours of the GLEAM 200 MHz image
        overlaid on the corresponding $60\arcsec$ TGSS image.

 
\section{Discussion}
   
    Combining the GLEAM and TGSS flux density measurements along
    with all available measurements in literature (see Table \ref{table:flux})
    we estimated the spectral indices of the detected haloes and relics
    in the range 80 -- 1400 MHz. In the case of
    PLCK G287.0+32.9, however, the spectrum is estimated over the range
    80 -- 3000 MHz.
    In Fig.~\ref{fig:spectra} we have shown the spectra 
    of the radio haloes and relics in the 9 clusters we observed.
    The spectra follow a power law over the frequency range $80-1400(/3000)$) MHz
    with no breaks.
    The mean values are $\alpha = -1.13\pm0.21$
    for haloes and $\alpha = -1.2\pm0.19$ for relics.
    These mean values are in agreement, within errors, with 
    the mean values of spectral indices in literature for 
    haloes ($\alpha = -1.34\pm 0.28$) and relics ($\alpha = -1.42\pm0.37$)
    \citep{feretti12}.
    Most of the spectral index estimates of haloes and relics in literature
    are in the range $325-1400$ MHz,
    while in the current study the estimates are in the range $80-1400$ MHz or
    $80-3000$ MHz.


    \begin{table*}
        \centering
        \begin{tabular}{llcccccccc}
        \hline
        Cluster	&	Object	&	z	&	$\alpha$	&	$\delta$	&	\multicolumn{3}{c}{Angular Size ($''\times'', ^{\circ}$)}   &   \multicolumn{2}{c}{Linear Size} \\
        &	& &	&	                                        &	$\theta_\text{maj}$	& $\theta_\text{min}$ & pa              &   kpc & kpc \\
        \hline
    A13		& R		& 0.094	& 00:13:28	& --19:29:58	& 150 & 108	&  36 	& 260	& 180	\\
    A548b	& B	& 0.042	& 05:45:22	& --25:47:07	&  84 &  72 &  45 	& 110	& 89	\\
            & A	& 		& 05:44:50	& --25:50:37	& 168 & 102 &  96 	& 180	& 130	\\
    A2063 	& R		& 0.035\\
    A2163 	& H		& 0.203	& 16:15:45	& --06:09:07	& 234 & 132 &  51 	& 970	& 390	\\
            & R		&		& 16:16:10	& --06:05:02	& 193 & 102	&  92 	& 660	& 350	\\
    A2254 	& H		& 0.178	& 17:17:52	& +19:40:37		& 147 & 109	&   0	& 450	& 340	\\
    A2345 	& R(E)	& 0.179	& 21:27:34	& --12:10:41	& 276 &  78	& 157	& 710	& 270	\\
            & R(W)	&		& 21:26:43	& --12:07:56	& 174 &  60	&  41 	& 590	& 180	\\
    A2744 	& H		& 0.308	& 00:14:20	& --30:23:29	& 192 & 132	&  63 	& 1010	& 690	\\
            & R		&		& 00:14:37	& --30:20:22	& 204 & 102 & 132	& 1100	& 400	\\
    PLCK G287.0+32.9& H		& 0.39	& 11:50:49	& --28:04:36	& 210 & 150 & 117	& 900	& 750	\\
            & R(SE)	&		& 11:51:12	& --28:11:41	& 252 &  96	&  77 	& 1530	& 670	\\
            & R(NW)	&		& 11:50:49	& --28:03:35	& 192 & 132	&  38 	& 860	& 410	\\
    RXC J1314.5-2515 & H		& 0.244\\
                & R(E)	& 	& 13:14:44	& -25:15:03	    & 192 & 120 & 143	& 660	& 600   \\
                & R(W)	& 	&\\
        \hline
        \end{tabular}
        \caption{Halos and relics detected by MWA.
        Columns 4 and 5 correspond to the peaks of emission at 200 MHz
        of the corresponding haloes and relics. No helo or relic
        was detected in A2063.
        The radio halo and the west relic in RXC J1314.5-2515
        are blended even in the 200 MHz GLEAM image.
        }
        \label{tab:angsize}
    \end{table*}

    \begin{figure*}
        \centering
        \includegraphics[width=0.7\linewidth,angle=270]{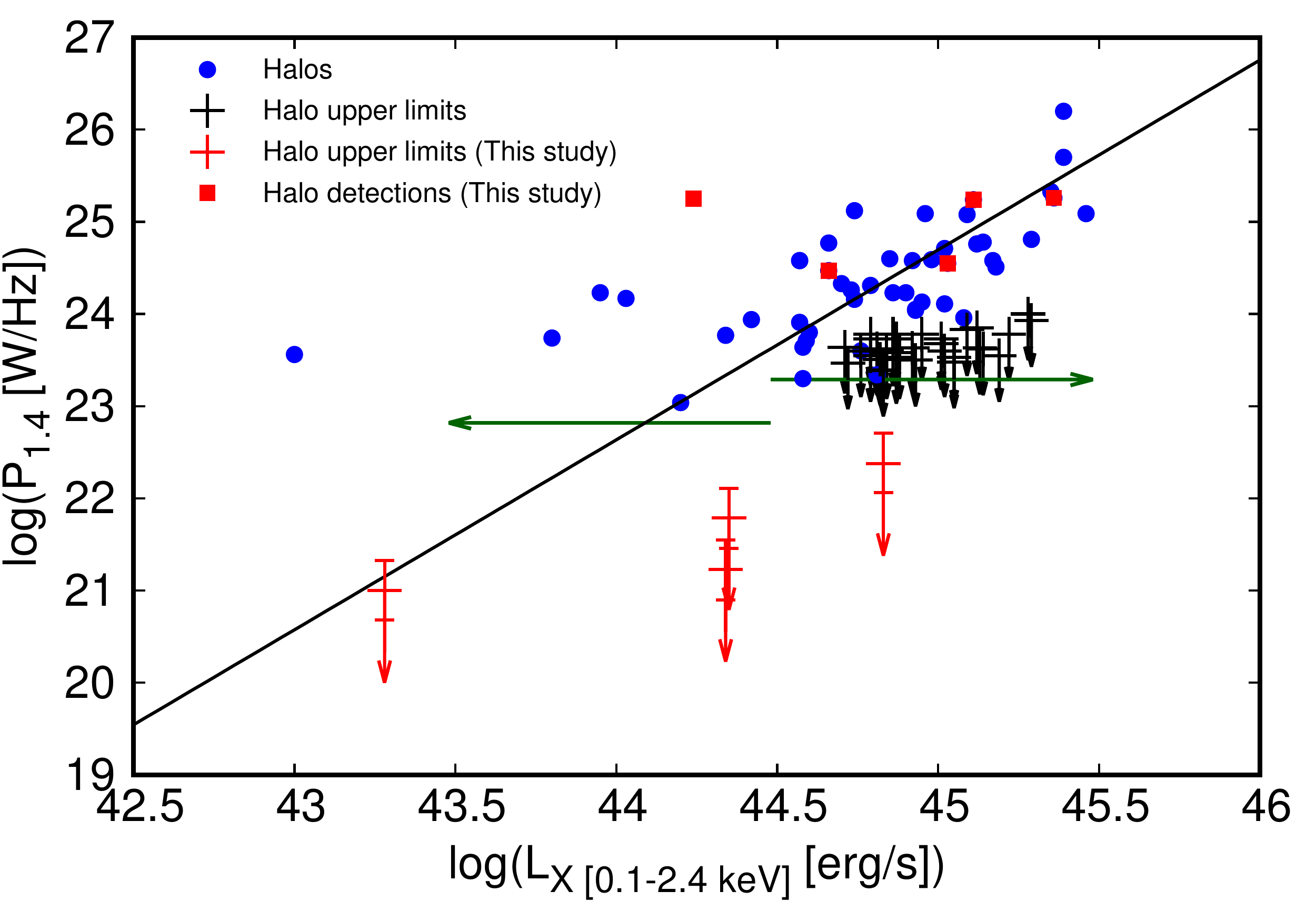}
        \caption{Plot showing the relation between the $L_{\rm X [0.1-2.4 keV]}$ of a cluster 
                and the radio power at 1.4 GHz ($P_{\rm 1.4}$) of the halo in the cluster.
                The filled blue circles represent all the known radio haloes
                \protect\citep{feretti12}.
                The solid black line is the best fit to the $L_{\rm X}-P_{\rm 1.4}$ relation 
                for radio haloes \protect\citep{brunetti09}.
                The black arrows represent the upper limits to halo emission
                \citep{venturi07,venturi08,kale13}.
                The MWA halo detections are shown as red squares while the upper limits 
                are shown as red arrows. Note that the MWA upper limits were estimated using 
                the RMS values estimated in the respective images at 200 MHz and 
                then extrapolating to 1.4 GHz with 
                a spectral index of $-1.34$.
                The four MWA upper limits, from left to right, are -- A548b, A2063, A13 and A2345.
                The five MWA halo detections, from left to right, are -- PLCK G287.0+32.9, A2254, 
                RXC J1314.4-2515, A2744 and A2163.
                The green arrows show the upper limits to the 
                radio powers in the \emph{off-state} radio haloes
                based on their X-ray luminosities as given by \protect\cite{brown11}.
                }
        \label{fig:lxp}
    \end{figure*}

    \begin{table}
        \centering
        \resizebox{\columnwidth}{!}{
        \begin{tabular}{lccc}
            \hline
            Cluster &   RMS     &   Resolution  & Linear Extent \\
                    &   (\mjyb) &   (\arcsec)   & (kpc)         \\
            \hline
            A13     &   9.2     &   133         & 240           \\
            A548b   &   7.8     &   132         & 120           \\
            A2063   &   19.2    &   121         & 88            \\
            A2345   &   9.2     &   136         & 420           \\
            \hline
        \end{tabular}
        }
        \caption{Upper limit clusters.
        The linear extents corresponding to the resolutions are given.}
        \label{tab200}
    \end{table}

    There are two exceptions to this result in our sample of clusters. 
    The first is A548b whose relics have flat spectral indices 
    ($-0.61$ (A) and $-0.38$ (B))
    and the other is A13 where the relic has a steeper  spectral index of $-1.74$.

    The steep spectrum of the A13 relic can be attributed to the fact that
    it is not a typical shock accelerated relic.
    X-ray analysis using {\it Chandra} \citep{juett08} shows that
    while the cluster is undergoing a merger event
    there are no shocks near the position of the radio relic.
    Moreover, the 1.4 GHz image of A13 shows this
    relic radio emission trailing
    from the brightest cluster galaxy located at the peak of the X-ray emission.
    The above authors suggest that
    this galaxy could have hosted an AGN in the past
    and the relic is a remnant of that emission.
    They propose two scenarios to explain this:
    either the radio plasma rose buoyantly to its current position,
    or the galaxy moved away during the merger event leaving behind
    the radio plasma that is now seen as a relic.
    In either scenario, there is no influx of new high energy electrons
    to this relic as the AGN is not active any more.
    As a result, the radio spectrum of this relic emission is steeper than the mean
    spectra of cluster relics which are produced by cluster merger shocks.
    This source is a relic radio galaxy rather than
    a cluster relic.

    In the case of A548b,
    it is debatable if the sources A and B are truly radio relics
    driven by merger shocks of the cluster
    or merely the lobes of radio emission produced by one or more of the
    cluster galaxies detected in the vicinity of these sources, including the source D (Fig. 1).


    It is possible to consider the sources A and B as relics in the cluster
    produced by merger shocks.
    There is some evidence supporting this scenario because
    the temperature distribution of the cluster shows a jump
    near the positions of the sources A and B \citep{solovyeva08}.
    However, as has been discussed in this paper, 
    the relics are toward the outskirts of the cluster and are located
    in a region of poor signal to noise ratio in the X-ray images.
    So, the evidence for the temperature jump near the relics is 
    ratehr marginal. Furthermore, the radio morphologies of the relics
    observed in the higher frequency images (\cite{feretti06}) do not
    show any sharp boundaries usually observed in other cluster relics, 
    but, are rather
    diffuse and irregular, resembling more like the radio emission 
    or, the relict radio emission produced by radio galaxies and / or quasars.

    It apears more likely that the sources A and B are lobes created by
    one or more of the cluster galaxies. In the case of source B, there is
    a small angular diameter ($4''\times 2''$) cluster radio source
    (ESO 488-G006) with a flat spectral index  embedded in this diffuse emission (\cite{feretti06}).
    In the case of source A, there are two cluster galaxies (source D and
    the extension seen toward south-west of source A in Fig. 1) within 
    $\sim$ 100 kpc of this diffuse radio emission
    either of which could be responsible for this diffuse
    radio emission.

     It appears that the sources A and B are more likely to be lobes of
     radio galaxies rather than cluster relics.


    Fig.~\ref{fig:lxp} shows 
    the empirical relation ($L_{\rm X}-P_{\rm 1.4}$)
    between the radio power of the halo at 1.4 GHz 
    ($P_{\rm 1.4}$) and the total X-ray luminosity of the cluster ($L_{\rm X}$) 
    in the range 0.1--2.4 keV.
    Based on the GMRT Radio Halo Survey,
    \cite{brunetti07,brunetti09} estimated upper limits 
    to the halo emission in clusters where none were detected.
    The upper limits plotted in Fig.~\ref{fig:lxp} (black crosses)
    \citep{venturi07,venturi08,kale13}
    correspond to clusters in the redshift range $z=0.2-0.4$
    and to an assumed halo size of 1 Mpc.
    These upper limits appear to correspond to \emph{off-state} clusters 
    compared to the detections which are in an \emph{on-state}.
    It has been known that, for the most part,
     \emph{on-state} clusters are the more disturbed clusters
    while the \emph{off-state} clusters are relaxed.
    However, if the \emph{hadronic} model is to be believed, there should be
    some amount of relativistic electrons in the ICM in either case.

    Five of the nine clusters studied in this paper contain radio haloes
    which were detected by  MWA.
    These clusters are shown in Fig.~\ref{fig:lxp}  with red squares.
    No radio haloes were detected in 
    the remaining four clusters -- A13, A548b, A2063 and A2345.
    Based on the RMS values in the respective images at 200 MHz,
    we have been able to put upper limits on the radio powers 
    of any possible radio haloes that might be present in these clusters.
    These RMS values were estimated after removing the already known non halo sources
    in the GLEAM 200 MHz images using BANE (see Section 2.1)
    and then estimating the RMS values from the central regions of the clusters
    in the residual images.
    The RMS values thus estimated correspond to radio powers of haloes
    whose linear extents are comparable to those of the spatial resolutions 
    in these images.
    The linear sizes corresponding to the spatial resolutions
    in these four images are given in column 4 of Table~\ref{tab200}.
    The upper limits shown in Fig.~\ref{fig:lxp} (red crosses)
    correspond to radio powers of such haloes.
    However, note that these upper limits at 1.4 GHz
    were obtained after extrapolating the respective values at 200 MHz
    with a spectral index of $-1.34$ which is the mean value
    of the spectral indices of haloes \citep{feretti12}.
    If the haloes in the current sample have spectra which are steeper than
    the mean value then the limits would be lower than those estimated here.
    The upper limit to the radio power of halo emission from A548b
    is consistent with that expected from the empirical
    $L_{\rm X}-P_{\rm 1.4}$ relation (best-fit line shown in Fig.~\ref{fig:lxp}).
    Note that this best-fit line \citep{brunetti09} corresponds
    to Giant Radio Haloes with linear extent, $l\gtrsim1$ Mpc.
    We discuss the upper limits on the remaining three clusters below.


    The upper limits on the radio powers of haloes at 1.4 GHz
    in A13, A2063 and A2345 (shown in Fig.~\ref{fig:lxp})
    are factors of $\sim$ 30, 120 and 90
    below that expected from the $L_{\rm X}-P_{\rm 1.4}$ relation respectively.
    If the extents of the radio haloes in these clusters
    are $\sim$ Mpc (canonical halo size) the upper limit obtained for A2063
    is consistent with the $L_{\rm X}-P_{\rm 1.4}$ relation.
    However the upper limits obtained for A13 and A2345
    are lower than the expected values by a factor of $\sim$ 2 and 20 respectively.

    Estimates of the upper limits on the radio powers of haloes in
    galaxy clusters have been carried out before. 
    \cite{brown11} used the radio continuum images of galaxy clusters from the
    Sydney University Molonglo Sky Survey (SUMSS, \citealt{bock99}) in order to estimate
    such constraints. 
    By stacking 105 clusters based on their X-ray luminosities,
    they found that for high X-ray luminosity
    ($L_{\rm X} > 3\times10^{44}$ ergs/s) clusters,
    the upper limit on the average radio power is
    $(1.95\pm0.75) \times10^{23}$ W Hz$^{-1}$,
    whereas for the low luminosity ($L_{\rm X} < 3\times10^{44}$ ergs/s)
    clusters, the upper limit on the average radio power is
    $(0.66\pm0.89)\times10^{23}$ W Hz$^{-1}$.
    The green arrows in Fig.~\ref{fig:lxp} show the upper limits
    for both low and high X-ray luminosity clusters.
    The individual upper limits obtained in this study with the MWA observations
    are comparable to, or,
    better than the average limits obtained by \cite{brown11}.

    It is instructive to discuss the upper limits on the radio powers
    of haloes in many clusters in the context of the hadronic, or, secondary model.        
    In order to estimate the contribution of the secondary model
    to radio emission from haloes, 
    it is necessary to first estimate the contribution of
    high energy cosmic ray (CR) protons to the energy content of the cluster.
    \cite{brunetti08} show (Fig. 3 in their paper)
    how the ratio of the relativistic protons
    to the thermal protons ($\epsilon_{\rm p}/\epsilon_{\rm th}$)
    varies as a function of the magnetic field ($B$)
    in the cluster.
    This relation depends on the spectral index ($\delta$)
    of the proton distribution
    ($N(p)\propto p^{-\delta}$, where $p$ is the particle momentum)
    as well as the temperature ($T$) and
    the number density of thermal protons ($n_{\rm th}$) in the cluster.
    Assuming $n_{\rm th}$ in the system to be $\sim 1500$ m$^{-3}$
    and the proton spectral index $\delta = 2.5$,
    Fig. 3 in \cite{brunetti08} shows how the energy density of
    cosmic-ray protons varies as a function of its magnetic field.
    Increasing the magnetic field from $0.5 \mu$G to $5 \mu$G 
    \citep{bonafede10}
    decreases the energy content of the cluster due to the CR proton component
    from 10\% to 0.25\%.
    Furthermore, \cite{brublas05} show that for a given magnetic field,
    the efficiency of electron acceleration decreases
    with increasing values of both $n_{\rm th}$
    as well as $\epsilon_{\rm p}/\epsilon_{\rm th}$, due to the increased damping
    of Alfv\'en waves on the protons.

    Recent analysis of the Coma cluster \citep{brulaz11}
    suggests that the strength of synchrotron emission due to the hadronic model
    is a factor of 10 smaller as compared to the primary model.
    The upper limits on the radio powers of haloes in A13 and A2345 are
    $\sim$ 10 times below the expected value. However, the upper limit on
    the radio power of a halo in A2063 is more than 50 times smaller
    than that expected.  It should be noted, however,
    that the number of unknowns in these systems is still quite large.
    While the electron density ($n_{\rm e}$),
    and hence the proton density ($n_{\rm p}$)
    in the clusters are known,
    the magnetic field, $B$,
    and the CR proton energy content, $\epsilon_{\rm p}/\epsilon_{\rm th}$,
    are still unknown.
    The cluster A2063, being a relaxed cluster could well
    host a weak magnetic field ($<$ 1 $\mu$G)
    resulting in a much weaker emission from the hadronic model.
    Independent measurements of magnetic fields in these
    clusters is crucial to the verification of the hadronic models.
    

\section{Summary}
    In this paper, we have presented low frequency observations of
    9 merging galaxy clusters using the Murchison Widefield Array.
    Theses observations were carried out at 5 frequencies
    (88, 118, 154, 188 and 215 MHz) as part of the 
    GaLactic and Extragalactic All-sky MWA (GLEAM) Survey.
    The images were products of a standard pipeline
    developed to analyse all of the GLEAM survey data.
    Furthermore, better resolution ($\sim$ 60$^{''}$)
    TGSS 150 MHz images were used to compare with the corresponding MWA images
    in order to identify and subtract unrelated sources
    that are blended with the halo and relic emission.
    
    We have detected radio relics in 7 of the 9 clusters
    and also estimated their spectra over the frequency range $80-1400$ MHz
    (or, in the case of PLCK G287.0+32.9, $80-3000$ MHz).
    These spectra were found to be fit by a power law over this range.
    The cluster A548b was found to contain radio sources
    (which were claimed to be relics)
    with the flattest spectral indices in the current sample ($-0.38$ and $-0.61$).
    Based on the current study we believe these sources to be radio lobes
    rather than relics produced in merger shocks.
    The relic in A13 was found to have the steepest spectrum in the current
    sample ($\alpha = -1.74$).
    However, the origin of this relic is not due to merger driven shocks
    but rather it is the remnant emission from an old radio galaxy.

    In 5 of the 9 clusters we detected the radio haloes
    that were first seen in high frequency ($\sim 1.4$ GHz) observations.
    Their spectra fit a power law with no breaks in the range $80-1400$ MHz.
    In the remaining 4 clusters where no radio haloes were detected,
    we have placed upper limits to their radio powers.
    These upper limits are a factor of $2-20$ below that expected from an 
    empirical relation between the X-ray luminosities of clusters
    and the radio powers of haloes in the corresponding clusters.
    Lack of independent measurements of magnetic fields in these clusters
    precludes one from putting serious constraints on the hadronic model at this
    stage.

\section*{Acknowledgement}
We thank the referee for detailed and critical comments.
This scientific work makes use of the Murchison Radio-astronomy Observatory,
operated by CSIRO.
We acknowledge the Wajarri Yamatji people
as the traditional owners of the Observatory site.
Support for the operation of the MWA
is provided by the Australian Government (NCRIS),
under a contract to Curtin University administered by Astronomy Australia Limited.
We acknowledge the Pawsey Supercomputing Centre
which is supported by the Western Australian and Australian Governments.
Parts of this research were conducted by the
Australian Research Council Centre of Excellence for All-sky Astrophysics
(CAASTRO), through project number CE110001020.
MJ-H, QZ and LH were supported in this work by grants to MJ-H
from the Marsden Fund and
the New Zealand Ministry of Business, Innovation and Employment.

\label{lastpage}

\bibliography{references}
\bibliographystyle{mnras}

\end{document}